\documentclass{IEEEtran}

\usepackage{float}
\usepackage{graphicx}
\usepackage{clrscode}
\usepackage{stfloats}
\usepackage{paralist}
\usepackage{tikz}
\usepackage{soul} 
\usepackage{color}
\usepackage{cite}
\usepackage{multicol}
\usepackage{comment}
\usepackage{placeins}

\usepackage{amsmath, amssymb, mathrsfs, amsfonts}
\usepackage{amsthm}
\usepackage{mathtools}
\usepackage{epsfig, epstopdf}

\usepackage{algorithm}
\usepackage{enumitem}
\usepackage{algorithmic}
\usepackage{lipsum}
\usepackage{pbox}
\usepackage{array}
\usepackage{tablefootnote}

\usepackage{subfigure}

\usetikzlibrary{arrows,automata}

\allowdisplaybreaks

\newtheorem{lemma}{Lemma}

\newtheorem{proposition}{Proposition}

\newtheorem{corollary}{Corollary}

\newtheorem{remark}{Remark}

\usepackage{hyperref}

\newcommand{\RNum}[1]{\uppercase\expandafter{\romannumeral #1\relax}}

\begin{document}
\title{Array Rotation in Near-Field ISAC:
Communication, Detection, and Estimation Under
Low-Complexity Transmit Architectures
}
\author{
		\IEEEauthorblockN{Marziyeh Soltani and Mark F. Flanagan, \textit{Senior Member, IEEE} \\
		\vspace*{0.5em}
			}\thanks{ This work was supported by Research Ireland under Grant 24/FFP-P/12895.\\
            Mark F. Flanagan and Marziyeh Soltani are with the School of Electrical and Electronic Engineering, University College Dublin, Belfield, Dublin 4, D04 V1W8, Ireland (e-mail: mark.flanagan@ieee.org, marziyeh.soltani@ucd.ie).}}

\maketitle
\begin{abstract}
This paper studies near-field integrated sensing and communication (ISAC)
with a rotatable base-station array under an exact spherical-wave propagation
model. For a general transmit covariance, we derive a Cram\'er--Rao bound
(CRB) decomposition that separates target-side geometry from
transmit-covariance-dependent terms. We then specialize the framework to
two representative low-complexity architectures: a single-stream superposition architecture, and a dual-waveform split architecture design. We characterize the communication-, detection-, and
estimation-optimal rotations for each architecture using the exact
finite-array model. In addition, we show how a second-order approximation yields compact approximate expressions for the optimal rotation angles. We
also study uncertainty in the target state used for beamforming design through a
correlated Gaussian perturbation model, and derive approximations of the optimal robust rotation
angles for target sensing and detection.
The analysis shows that the two architectures behave differently under
rotation: the single-stream design
can align communication and detection tasks, whereas the dual-waveform design exhibits a direct communication--detection tradeoff and can incur a larger
communication loss when sensing is prioritized. Numerical results show clear
gains over a fixed array orientation, especially for range estimation, and close
agreement between the approximate and exact rotation choices. Detection is
less affected by target-state errors than estimation in the considered
cases.
\end{abstract}
\section{Introduction}\label{introduction}
Integrated sensing and communication (ISAC) is a promising paradigm for future wireless systems as it provides a common platform and spectrum resource that can support both data transmission and environmental awareness \cite{Liu22,Lu24}. With large apertures and high carrier frequencies, many users and targets are situated in the radiative near-field, where the planar-wave model is no longer adequate and the array response depends jointly on angle, range, and aperture geometry \cite{Lu21b,Wan23,Wan23c,Hua24}. In this regime, both communication and target sensing (i.e., target detection or parameter estimation) are governed by the same spherical-wave structure.

Most existing works on near-field ISAC assume a fixed array orientation and optimize the transmit covariance, beamformer, and power allocation \cite{Wan23c,Hua24,Li23b,Zha23,Luo25,Ray26}. More recently, rotatable arrays have been recognized as a new geometric degree of freedom for ISAC \cite{Zho25,Zha26,Sun25}. Rotation changes both the angular alignment and the
wavefront-curvature mismatch between the user and the target, thereby
creating rotation-dependent communication--sensing tradeoffs beyond
conventional beamforming.
However, the role of rotation in near-field ISAC is still not well
understood analytically, especially when communication and sensing are
coupled within structured transmit architectures
\cite{Zho25,Zha26}. It remains unclear when rotation is cooperative, with
communication and sensing favoring the same or nearby orientations, and when
it becomes a genuine shared-resource tradeoff.

Motivated by this gap, in this paper we study how array rotation affects two
representative low-complexity transmit architectures for near-field ISAC. 
Specifically, we consider two architectures: a single-stream (SS)
superposition architecture and a dual-waveform (DW) split
architecture.
Rather than redesigning a general beamformer at every rotation angle, we
keep the transmit structures prescribed and characterize how rotation changes
their communication, detection, and estimation performance. This viewpoint
separates the effects of rotation on target illumination, target-side
range--angle geometry, and the way the received signal changes with target
range and angle under the chosen transmit structure. It therefore reveals
why the two architectures can favor different rotations and whether
communication, detection, and estimation can benefit from the same
orientation or instead impose a tradeoff. A numerical joint covariance-and-rotation optimizer can provide operating
points, but it does not directly separate these underlying mechanisms.
The prescribed structures also
retain physically interpretable ISAC designs that include communication-only and sensing-only transmission as special cases. This focus is consistent with prior ISAC studies that analyze sensing
and communication performance under structured beamforming designs
\cite{Ngu24b,Shi26b,Wu26f,Ouy23b,Zha23}.
\subsection{Contributions}
We study a narrowband monostatic near-field ISAC system in which a base station (BS) equipped with a rotatable antenna array simultaneously serves one communication user and senses one target under exact spherical-wave phase propagation while neglecting
path-loss variation across the aperture. The main contributions of this paper are summarized as follows.

\noindent $\bullet$ For an arbitrary transmit covariance, we derive an exact
range--angle Fisher-information matrix (FIM) decomposition that separates target-side
range--angle geometry from transmit-covariance-dependent information terms,
making explicit how array rotation and transmit structure jointly affect
range--angle estimation.

\noindent $\bullet$ We show that at every rotation and under the same user QoS constraint,
(i) the CRB-optimal rank-one beamformer lies in the SS family, and
(ii) the
detection-optimal transmit covariance among all covariances is attained by an
SS beamformer. Consequently, no DW design meeting the same user SINR requirement can
outperform SS for detection.

\noindent $\bullet$ We derive the user signal-to-interference-plus-noise ratio (SINR), range–angle Cramér–Rao bounds (CRBs), and the detection probability under the location-conditioned generalized likelihood ratio test (GLRT). For prescribed SS beamforming coefficients and DW power fractions,
we formulate the corresponding exact finite-array communication-, detection-,
and estimation-centric rotation problems, derive their stationary-point conditions, and
use a second-order approximation to obtain compact rotation
objectives. In selected geometries, these yield closed-form
approximate or globally optimal rotation angles.

\noindent $\bullet$ We show that the communication- and detection-optimal rotations can
coincide in the case of the SS architecture, whereas they conflict in the case of DW when the sensing stream acts as interference at the user. In the considered cases, the estimation-optimal
rotation causes only a modest communication loss in SS, whereas in DW it
causes a smaller detection-information loss than in SS but a substantially larger
communication loss.

\noindent $\bullet$ We extend the framework to the case where imperfect target state knowledge is modeled using a correlated Gaussian perturbation model and we derive
low-complexity approximations for robust sensing- and detection-centric
rotations. Numerical results show that the approximate rotations closely
match the exact optima, that rotation provides clear gains over
a fixed orientation, most notably for range estimation, and that detection is
less sensitive to target-state errors than estimation. We also compare SS and DW with the joint transmit-covariance and array-rotation design of \cite{Zha26}, specialized to the single-user, single-target case and evaluated using the same trace-CRB objective. For the tested geometry and SINR range, SS remains close to this
trace-CRB reference, whereas DW shows a larger CRB gap.
\subsection{Existing work}
\noindent\textbf{Fixed-orientation near-field ISAC:}
Recent near-field ISAC studies have investigated beam focusing and
communication--sensing tradeoffs under spherical-wave propagation, usually
with a fixed array orientation and optimized covariance, beamforming, or
power allocation \cite{Wan23c,Hua24,Li23b,Zha23,Luo25,Ray26}. In particular,
\cite{Zha23} develops communication-centric, sensing-centric, and
Pareto-optimal beamforming designs for a fixed orientation.

\noindent\textbf{Near-field sensing CRB analysis:}
Exact and asymptotic range--angle CRB analyses have been developed for
near-field sensing under spherical-wave propagation
\cite{Wan23,Wan24,Jia24c}. These works provide important foundations for
near-field estimation, primarily from a sensing-only transmission perspective.

\noindent\textbf{Rotatable and movable arrays for ISAC:}
Recent studies investigate rotatable or movable arrays for near-field ISAC
using covariance, position, rotation, or secrecy-rate optimization
\cite{Zho25,Zha26,Sun25,Wan25,Qu25,Wan26o}. The work \cite{Zha26}, which is closest to that of this paper, considers a broader multi-user, single-target framework with
joint general-covariance and rotation design. It derives closed-form
rotations for communication-only and sensing-only transmission, while solving the
general mixed design problem numerically.
\subsection{Comparison to Existing Work}
In contrast to fixed-orientation ISAC such as \cite{Zha23}, in this paper rotation is an explicit design variable, and sensing is characterized through both
range--angle CRBs and location-conditioned GLRT detection rather than
through sensing rate alone. The most closely related work, \cite{Zha26}, jointly optimizes a general
transmit covariance and the array rotation for a multi-user, single-target
system, deriving closed-form rotations for communication-only and
sensing-only transmission and solving the mixed design numerically. Our work is complementary: it explains how array rotation shapes communication, detection, and estimation tradeoffs under low-complexity structured transmit architectures, and provides practical design insights.
we study a single-user, single-target system with
prescribed transmit structures, which enables an exact finite-array
analytical characterization of why and how rotation affects each task.
Specifically, compared with \cite{Zha26} and fixed-orientation studies such
as \cite{Zha23}, this paper
(i) derives an exact FIM decomposition for an arbitrary transmit covariance, separating target-side geometry from transmit-covariance effects and showing how each shapes range–angle estimation accuracy, then specializes the result to the SS and DW architectures;
(ii) proves that SS is optimal for detection among all transmit covariances
at every rotation, and hence dominates DW for detection;
(iii) characterizes, for each architecture, when the communication\mbox{-,}
detection-, and estimation-optimal rotations coincide and when they conflict,
including bounds on the resulting losses;
(iv) considers location-conditioned GLRT detection in addition to the CRB; and
(v) derives robust rotation angles under correlated target-state
perturbations.
The communication- and sensing-centric designs of \cite{Zha23} and the
broadside-optimality result of \cite{Zha26} appear as special cases of our
framework.

The remainder of this paper is organized as follows. Section~\ref{systemmodel}
presents the system model and problem formulation. Sections~\ref{sensingestimation}
and~\ref{detectionanalysis} analyze sensing from the estimation and detection
perspectives, respectively. Section~\ref{approximations} develops the
analytical approximations. Sections~\ref{Communicationcentricrotation},
\ref{comdetcompat}, and~\ref{sensingcentric} analyze communication-centric
rotation optimization, communication--detection synergy, and sensing-centric rotation optimization,
respectively. Section~\ref{specialcases} presents special cases with explicit optimal rotations, while Section~\ref{robustness} develops robust rotation designs under target-state perturbations. Section~\ref{numerical} presents the numerical results, and Section~\ref{conclusion} concludes the paper.

\textbf{Notation:} Bold uppercase and lowercase letters represent matrices and column vectors, respectively. The sets of real and complex numbers are denoted by \(\mathbb{R}\) and \(\mathbb{C}\). The operators \((\cdot)^\top\), \((\cdot)^*\), and \((\cdot)^H\) represent transpose, conjugate, and Hermitian transpose, respectively. \(\operatorname{tr}(\cdot)\), \(\operatorname{det}(\cdot)\), and \(\mathbb{E}[\cdot]\) denote the trace, determinant, and expectation, respectively; \(\mathcal{CN}(\mathbf{0},\mathbf{\Sigma})\) denotes a zero-mean circularly symmetric complex Gaussian distribution with covariance matrix \(\mathbf{\Sigma}\); and \(\mathbf{I}_N\) is the \(N\times N\) identity matrix.
\section{System Model and Problem Formulation}\label{systemmodel}
\begin{figure}[!t]
\centering
\includegraphics[scale=.44]{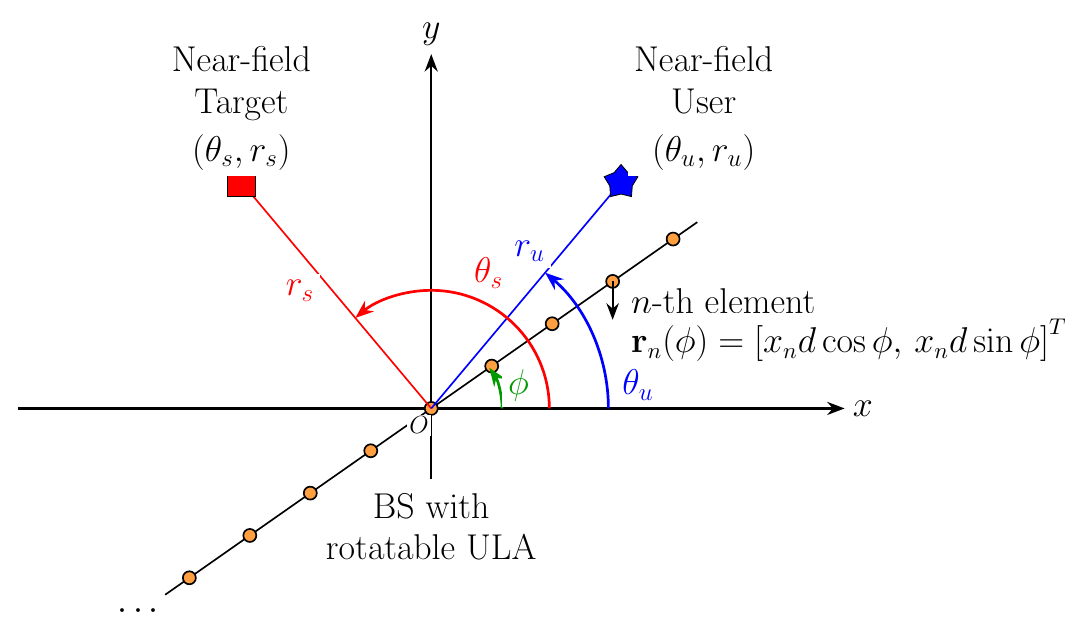}
\caption{The investigated near-field ISAC system model with rotatable array.}
\label{model}
\end{figure}

We consider a narrowband monostatic near-field ISAC system with one communication user and one sensing target, as illustrated in Fig.\ref{model}. The BS employs an $N$-antenna rotatable uniform linear array (ULA) with inter-element spacing $d=\lambda/2$, where $\lambda$ is the carrier wavelength. The array aperture is $D=(N-1)d$, and its orientation is controlled by a one-dimensional rotation angle $\phi$.
\subsection{Channel Models}\label{channelmodel}
Without loss of generality, we assume a two-dimensional coordinate system in which the center of the ULA is located at the origin. In the local array coordinate system before rotation, the \(n\)-th antenna element, $n\in\{1,\ldots,N\}$, is located at
$[\chi_n d,0]^\top,$ where $\chi_n=n-\frac{N+1}{2}.$ After rotating the ULA by the angle \(\phi\), the corresponding coordinates in the global coordinate system become  $\mathbf r_n(\phi)=[\chi_n d\cos\phi,\chi_n d\sin\phi]^\top.$
For a point located at angle $\theta$ and range $r$ relative to the array center, the exact spherical-wave distance from the $n$-th antenna element is
$
r_n(\theta,r,\phi)
=
\sqrt{r^2+\chi_n^2d^2-2r\chi_nd\cos(\theta-\phi)}.
$
The corresponding near-field steering-vector element is
$
b_n(\theta,r,\phi)
=
e^{-jk(r_n(\theta,r,\phi)-r)},
\qquad
k=\frac{2\pi}{\lambda},
$
and the complete near-field steering vector is
$
\mathbf b(\theta,r,\phi)
=
[b_1(\theta,r,\phi),\ldots,b_N(\theta,r,\phi)]^\top.
$
The steering vectors for the communication user and the sensing target are thus
$\mathbf b_u(\phi)\triangleq \mathbf b(\theta_u,r_u,\phi)$ and
$\mathbf b_s(\phi)\triangleq \mathbf b(\theta_s,r_s,\phi),$
respectively.
We assume that both lie in the radiative near-field region, i.e.,
$
0.62\sqrt{\frac{D^3}{\lambda}}
\lesssim r_v \lesssim
\frac{2D^2}{\lambda},$ for
$v\in\{u,s\}$ \cite{selvan}.
The communication channel is modeled as a line-of-sight link:
$
\mathbf h_u(\phi)=\beta_u\mathbf b_u,
$
where
$
\beta_u=\sqrt{\zeta r_u^{-\alpha}},
\qquad
\zeta=\left(\frac{\lambda}{4\pi}\right)^2,
$
and $\alpha$ is the path-loss exponent.
We assume that the sensing signal is received by the same $N$-antenna
ULA used for transmission. Accordingly, the monostatic sensing channel is
modeled as\footnote{Unless needed for emphasis, we will suppress the argument
$(\phi)$ in rotation-dependent quantities and write, for example,
$b_u$, $b_s$, $v_u$, $v_s$, $G$, and $\rho$ in place of
$b_u(\phi)$, $b_s(\phi)$, $v_u(\phi)$, $v_s(\phi)$, $G(\phi)$, and
$\rho(\phi)$ when the dependence is clear from the context. However, we
will make the argument explicit when optimizing over $\phi$ or when
additional variables, such as perturbations, are present.}
$
\mathbf H_s(\phi)
=
\beta_s\mathbf b_s(\phi)\mathbf b_s^H(\phi)
\in\mathbb C^{N\times N},
$
where $\beta_s\in\mathbb C$ denotes the unknown effective complex
round-trip coefficient which absorbs the effects of deterministic attenuation and target
reflectivity.
We assume equal path loss across the array aperture, i.e.,
$
D/r_u\ll1$ and $D/r_s\ll1.
$
Thus, the model retains exact spherical-wave phase propagation while
neglecting element-wise path-loss variation across the aperture. The antenna
elements are assumed to be isotropic, so rotation changes the array geometry but
not the element radiation pattern\footnote{Directivity, polarization, mutual
coupling, and pattern-dependent rotation effects are outside the scope of the present study.}.
Define the normalized steering vectors
$
\mathbf v_u=\frac{\mathbf b_u}{\sqrt N},
\qquad
\mathbf v_s=\frac{\mathbf b_s}{\sqrt N}.
$
By defining $\Delta r_{v,n}\triangleq r_n(\theta_v,r_v,\phi)-r_v$,\quad $v\in\{u,s\}$, the \textit{user-target steering correlation} can be defined as $\rho(\phi)
=
\mathbf v_u^H\mathbf v_s
=
\frac{1}{N}\sum_{n=1}^{N} b_{u,n}^*b_{s,n}=
\frac{1}{N}\sum_{n=1}^{N}
e^{-j\frac{2\pi}{\lambda}\left(\Delta r_{s,n}-\Delta r_{u,n}\right)}$; as we shall see, this plays a central role in the communication--sensing coupling studied in this paper.
\subsection{Transmitted Signal}\label{transmittedsignal}
The BS transmits a narrowband dual-functional radar-communication signal $\mathbf X\in\mathbb C^{N\times T}$, where $T$ is the block length. We consider two transmit architectures: a single-stream superposition architecture (SS) and a dual-waveform split architecture (DW). In the following, superscripts $(S)$ and $(D)$ denote the SS and DW architectures, respectively.
\\
\textbf{1) SS architecture:}
The first architecture uses a rank-one beamformer in the span of the user and target steering vectors:
\begin{align}
\mathbf w^{(S)}(c_u,c_s,\phi)
=
\frac{c_u\mathbf v_u(\phi)+c_s\mathbf v_s(\phi)}
{\sqrt{Z^{(S)}(c_u,c_s,\phi)}},
\qquad
(c_u,c_s)\neq(0,0),
\label{beamA}
\end{align}
where $c_u,c_s\in\mathbb C$ are combining coefficients and
$
Z^{(S)}(c_u,c_s,\phi)
=
\|c_u\mathbf v_u(\phi)+c_s\mathbf v_s(\phi)\|^2
=
|c_u|^2+|c_s|^2+2\Re\{c_u^*c_s\rho(\phi)\}>0.
$ The normalization by $Z^{(S)}(c_u,c_s,\phi)$ ensures that $\|\mathbf w^{(S)}(c_u,c_s,\phi)\|^2=1$. The transmitted signal matrix is
$
\mathbf X
=
\sqrt{P_t}\,\mathbf w^{(S)}(c_u,c_s,\phi)\mathbf s^\top,
$
where $P_t$ is the total transmit power. Here, $\mathbf{s}\in\mathbb{C}^{T\times1}$ contains i.i.d.
$\mathcal{CN}(0,1)$ symbols. Hence, for sufficiently large $T$,
$
\frac{1}{T}\mathbf{s}^H\mathbf{s}\approx1$ \cite{CrameRaoBoundOptimizationforJoint}.
The normalized sample covariance, $\mathbf R_x
\triangleq \frac{1}{TP_t}\mathbf X\mathbf X^H$, is therefore approximated by
$\mathbf R_x=
\mathbf w^{(S)}\mathbf w^{(S)H},$
where $\operatorname{tr}(\mathbf R_x)=1$.
\\
\textbf{2) DW architecture:}
The DW architecture uses separate communication and sensing waveforms. The
transmitted signal matrix is
\begin{align}
\mathbf X
=
\sqrt{P_u}\mathbf v_u\mathbf s_u^\top
+
\sqrt{P_s}\mathbf v_s\mathbf s_s^\top ,
\end{align}
where $P_u=\mu_uP_t$, $P_s=\mu_sP_t$, and
$\mu_u+\mu_s=1$. The communication signal
$\mathbf s_u\in\mathbb C^{T\times1}$ contains i.i.d.
$\mathcal{CN}(0,1)$ symbols, while the fixed sensing waveform
$\mathbf s_s\in\mathbb C^{T\times1}$ satisfies
$
\frac{1}{T}\mathbf s_s^H\mathbf s_s=1.
$
For sufficiently large $T$, the law of large numbers gives
$
\frac{1}{T}\mathbf{s}_u^H\mathbf{s}_u\approx1$ and $\frac{1}{T}\mathbf{s}_u^\top\mathbf{s}_s^*\approx0.
$
Therefore, the normalized sample covariance is approximated by
$\mathbf R_x
=
\mu_u\mathbf v_u\mathbf v_u^H
+
\mu_s\mathbf v_s\mathbf v_s^H,$
with $\operatorname{tr}(\mathbf R_x)=\mu_u+\mu_s=1$.
\begin{remark}
The SS architecture is motivated by two results which will be derived later. For a fixed
rotation angle $\phi$, Corollary~\ref{procrbrankoneproofedition} shows that
the CRB-optimal rank-one beamformer under a user-QoS constraint lies in
$\mathrm{span}\{\mathbf b_u(\phi),\mathbf b_s(\phi)\}$, and
Lemma~\ref{thm:detection-optimality} shows that, among all transmit
covariances satisfying the same QoS constraint, the detection optimum is
attained by a beamformer in the same span. The DW architecture is the
minimal extension that introduces a dedicated sensing waveform and hence a
rank-two covariance structure. Both architectures include
communication-only and sensing-only transmission as special cases,
obtained by setting one coefficient or power fraction to zero.
\end{remark}
\subsection{Problem Formulation and Metrics}\label{problemformulation}
We study target sensing from two complementary viewpoints: detection and estimation. In the main analysis, we assume that the target state \((r_s,\theta_s)\) is perfectly known at the transmitter through a previous sensing, tracking, or coherence block. The current waveform, beamformer, and rotation angle are then designed for this state \cite{Hua22,Hua24,Qu23,Zha26}. The analysis is therefore conditional on a specified target state, which is consistent with the standard local interpretation of deterministic-parameter CRB analysis \cite{Wan23}. Under this viewpoint, the CRB quantifies local range--angle refinement accuracy, whereas the detection metric measures target confirmation for the
same tracked state \cite{Jay25,Gal24}. This distinction is especially relevant in the near field, where sensing is intrinsically two-dimensional because of spherical-wave propagation \cite{Wan23,Wan23c,Qu23}. In Section~\ref{robustness}, we will relax the perfect-state-information assumption by allowing the transmitter-side target state information to be imperfect and model the mismatch between the design state $(\hat r_s,\hat\theta_s)$ and the true target state using a correlated Gaussian perturbation model.

Because rotating a centered symmetric ULA by $\pi$ only reverses the order
of its elements, the communication, detection, and estimation metrics
considered here are unchanged. Hence, without loss of generality, we restrict
the rotation angle to
$
\Phi\triangleq[-\pi/2,\pi/2).
$ The selected rotation is held fixed over each tracking or coherence interval
and updated when the target-state estimate is refreshed.
Let
$
\vartheta_S=(c_u,c_s),
\qquad
\vartheta_D=\mu_u.
$
For SS, define the communication-, estimation-, and detection-centric
rotations by
$
\phi_u^{(S),\star}(\vartheta_S)
=
\arg\max_{\phi\in\Phi}
\mathrm{SNR}_u^{(S)}(\vartheta_S,\phi),
$
$
\phi_s^{(S),\star}(\vartheta_S)
=
\arg\min_{\phi\in\Phi}
\log\det\!\big(
\mathrm{CRB}^{(S)}(\vartheta_S,\phi)
\big),
$
and
$
\phi_d^{(S),\star}(\vartheta_S)
=
\arg\max_{\phi\in\Phi}
P_D^{(S)}(\vartheta_S,\phi).
$
For DW, define
$
\phi_u^{(D),\star}(\vartheta_D)
=
\arg\max_{\phi\in\Phi}
\mathrm{SINR}_u^{(D)}(\vartheta_D,\phi),
$
$
\phi_s^{(D),\star}(\vartheta_D)
=
\arg\min_{\phi\in\Phi}
\log\det\!\big(
\mathrm{CRB}^{(D)}(\vartheta_D,\phi)
\big),
$
and
$
\phi_d^{(D),\star}(\vartheta_D)
=
\arg\max_{\phi\in\Phi}
P_D^{(D)}(\vartheta_D,\phi).
$ Here, $P_D^{(A)}(\vartheta_A,\phi)$ denotes the target detection
probability under architecture $A\in\{S,D\}$, while
$\operatorname{CRB}^{(A)}(\vartheta_A,\phi)$ denotes the joint
range--angle CRB matrix for $(r_s,\theta_s)$.

We will derive the communication metrics under the two transmit architectures, followed by the sensing metrics from estimation and detection viewpoints. Let $\mathbf X[t]\in\mathbb C^{N\times 1}$ denote the $t$-th column of $\mathbf X$. The received signal at the communication user at time slot $t$ is
$
y_u[t]
=
\mathbf h_u^H\mathbf X[t]+n_u[t]
=
\beta_u\mathbf b_u^H\mathbf X[t]+n_u[t],$ where $
n_u[t]\sim\mathcal{CN}(0,\sigma_u^2).
$
Using
$
\mathbf b_u^H\mathbf v_u=\sqrt N,
\qquad
\mathbf b_u^H\mathbf v_s=\sqrt N\,\rho(\phi),
$
the user signal-to-noise ratio (SNR) under SS is
\begin{align}
\mathrm{SNR}_u^{(S)}(c_u,c_s,\phi)
=
\frac{P_t|\beta_u|^2N}{\sigma_u^2}
\frac{|c_u+c_s\rho(\phi)|^2}
{|c_u|^2+|c_s|^2+2\Re\{c_u^*c_s\rho(\phi)\}}.
\label{SINRA}
\end{align}
Under DW, treating the user-directed stream as desired and the sensing stream as interference\footnote{This model assumes that the user does not have the information
needed to cancel the sensing waveform and therefore treats it as noise.
With ideal cancellation, the user SNR becomes
$\mathrm{SNR}_{u,\mathrm{cancel}}^{(D)}
=\mu_uP_t|\beta_u|^2N/\sigma_u^2$, which is independent of $\phi$.
Thus, the communication--detection rotation conflict in
Section~\ref{comdetcompat} is specific to the no-cancellation model,
while all estimation and detection metrics are unchanged.}, the user SINR is
\begin{align}
\mathrm{SINR}_u^{(D)}(\mu_u,\phi)
=
\frac{\mu_uP_t|\beta_u|^2N}
{(1-\mu_u)P_t|\beta_u|^2N|\rho(\phi)|^2+\sigma_u^2}.
\label{SINRB}
\end{align}
\section{Sensing Performance Analysis: Estimation}\label{sensingestimation}
In this section we present the performance analysis of range-angle estimation in four steps.
We first derive the FIM for joint range–angle estimation under an arbitrary transmit covariance. We then decompose this FIM into target-side spherical-wave geometry terms and transmit-covariance-dependent terms that capture target illumination and how the received echo changes with target range and angle. We then specialize this decomposition to the rank-one case, which contains the SS beamformer as a special case. Finally, we derive the
CRB for DW and compare the FIMs for the SS and DW architectures.

The received sensing signal is
$
\mathbf Y_s
=
\mathbf H_s\mathbf X+\mathbf Z_s
\in\mathbb C^{N\times T}.
$
Here, $\mathbf Z_s$ denotes thermal receiver noise\footnote{We assume that the target’s round-trip delay is sufficiently long to prevent interference between the transmitted signal and the received echo.}, modeled as
$
\operatorname{vec}(\mathbf Z_s)
\sim\mathcal{CN}(\mathbf 0,\sigma_r^2\mathbf I).
$
After vectorization,
$
\mathbf y
=
\operatorname{vec}(\mathbf Y_s)
=
\beta_s\mathbf u+\mathbf z,
$
where
$
\mathbf u(\theta_s,r_s)
=
\operatorname{vec}
\big(\mathbf b_s\mathbf b_s^H\mathbf X\big),
\qquad
\mathbf z=\operatorname{vec}(\mathbf Z_s).
$
The unknown parameter vector is\footnote{The parameter $\beta_s$ is an independent unknown
complex nuisance parameter and is held fixed when differentiating with respect to $r_s$ and
$\theta_s$, consistent with existing near-field CRB analyses
\cite{Wan23,Qu23,Zha26}.}
$
\boldsymbol\xi
=
[\nu^\top,\boldsymbol{\epsilon}^\top]^\top,$ where $
\nu=[r_s,\theta_s]^\top$ and $\boldsymbol{\epsilon}
=
[\Re\{\beta_s\},\Im\{\beta_s\}]^\top.
$
Define
$
\mathbf G(\phi)
\triangleq
\mathbf b_s(\phi)\mathbf b_s^H(\phi),$ and
$T_0
\triangleq
\operatorname{tr}\!\big(
\mathbf G\mathbf R_x\mathbf G^H
\big).
$
For $\iota,\jmath\in\{r,\theta\}$, define
$
\dot{\mathbf G}_{\iota}(\phi)
\triangleq
\frac{\partial\mathbf G(\phi)}{\partial\iota},
\qquad
T_{\iota}
\triangleq
\operatorname{tr}\!\big(
\mathbf G\mathbf R_x\dot{\mathbf G}_{\iota}
\big),
$
and
$
T_{\iota\jmath}
\triangleq
\operatorname{tr}\!\big(
\dot{\mathbf G}_{\iota}
\mathbf R_x
\dot{\mathbf G}_{\jmath}
\big).
$ The following proposition gives the FIM and CRB for an arbitrary
transmit covariance.
\begin{proposition}
\label{generalCRBprop}
The FIM for
$
\boldsymbol{\nu}=[r_s,\theta_s]^\top,
$ is
\begin{align}
\mathbf J
=
\frac{2|\beta_s|^2TP_t}{\sigma_r^2T_0}
\begin{bmatrix}
m_{11}&m_{12}\\
m_{12}&m_{22}
\end{bmatrix},
\end{align}
where
$
m_{11}=T_{rr}T_0-|T_r|^2,
\quad
m_{22}=T_{\theta\theta}T_0-|T_\theta|^2,
\quad
m_{12}=\Re\{T_{\theta r}T_0-T_rT_\theta^*\}.
$
Consequently, $\operatorname{CRB}(\nu)=\mathbf J^{-1}$, and the
range and angle CRBs are
$
\operatorname{CRB}_{r_s}
=
\frac{\sigma_r^2}{2|\beta_s|^2TP_t}
\frac{T_0m_{22}}{m_{11}m_{22}-m_{12}^2}
$
and
$
\operatorname{CRB}_{\theta_s}
=
\frac{\sigma_r^2}{2|\beta_s|^2TP_t}
\frac{T_0m_{11}}{m_{11}m_{22}-m_{12}^2},
$
respectively.
\end{proposition}
The proof of Proposition~\ref{generalCRBprop} is given in
Appendix~\ref{app:generalcrb}\footnote{We restrict the CRB optimization to rotations for which the FIM for $(r_s,\theta_s)$ is positive definite, so that
$\operatorname{CRB}(\nu)=\mathbf J^{-1}$ is well defined.}.

Let
$
\Delta_s\triangleq\theta_s-\phi,
\qquad
a_n\triangleq
\frac{\partial\Delta r_{s,n}}{\partial\theta_s},
\qquad
c_n\triangleq
\frac{\partial\Delta r_{s,n}}{\partial r_s}.
$
Define the geometry-related sums
$
S_{a,2}=\sum_{n=1}^N a_n^2,
\qquad
S_{c,2}=\sum_{n=1}^N c_n^2,
\qquad
S_{a,c}=\sum_{n=1}^N a_nc_n.
$
Define the geometry-related means
$
\bar a\triangleq\frac{\sum_{n=1}^N a_n}{N},
\qquad
\bar c\triangleq\frac{\sum_{n=1}^N c_n}{N},
$
and the centered geometry-related moments
$
V_\theta
=
\frac{S_{a,2}}{N}-\bar a^2,
\qquad
V_r
=
\frac{S_{c,2}}{N}-\bar c^2,
\qquad
C_{r\theta}
=
\frac{S_{a,c}}{N}-\bar a\bar c.
$
For the transmit-covariance-dependent terms, define the target steering
derivatives
$
\mathbf b_{s,r}
\triangleq
\frac{\partial\mathbf b_s}{\partial r_s},
\qquad
\mathbf b_{s,\theta}
\triangleq
\frac{\partial\mathbf b_s}{\partial\theta_s},
$
and their components orthogonal to $\mathbf b_s$:
$
\widetilde{\mathbf b}_r
\triangleq
\mathbf b_{s,r}
-
\frac{\mathbf b_s^H\mathbf b_{s,r}}{N}\mathbf b_s,
\qquad
\widetilde{\mathbf b}_\theta
\triangleq
\mathbf b_{s,\theta}
-
\frac{\mathbf b_s^H\mathbf b_{s,\theta}}{N}\mathbf b_s.
$ The following lemma derives the geometry--covariance decomposition of CRBs used
to analyze both architectures.
\begin{lemma}\label{geometryCRBlemma}
For an arbitrary transmit covariance matrix $\mathbf R_x$, the range and angle CRBs can be decomposed as
$
\operatorname{CRB}_{r_s}
=\frac{\sigma_r^2}{2|\beta_s|^2TP_t}
\frac{g_0A_\theta}
{N(A_rA_\theta-A_c^2)}$ and $\operatorname{CRB}_{\theta_s}
=
\frac{\sigma_r^2}{2|\beta_s|^2TP_t}
\frac{g_0A_r}
{N(A_rA_\theta-A_c^2)}.
$
Moreover,
$
\det\!\big(\operatorname{CRB}(\nu)\big)
=
\left(\frac{\sigma_r^2}{2|\beta_s|^2TP_t}\right)^2
\frac{g_0^2}
{N^2(A_rA_\theta-A_c^2)}.
$
where
$
A_r
=
g_0h_{rr}
+
k^2g_0^2V_r
-
|g_1|^2,
$
$
A_\theta
=
g_0h_{\theta\theta}
+
k^2g_0^2V_\theta
-
|g_2|^2,
$
$
A_c
=
g_0\Re\{h_{\theta r}\}
+
k^2g_0^2C_{r\theta}
-
\Re\{g_1g_2^*\},
$
with
$
g_0
=
\mathbf b_s^H\mathbf R_x\mathbf b_s,
\qquad
g_1
=
\mathbf b_s^H\mathbf R_x\widetilde{\mathbf b}_r,
\qquad
g_2
=
\mathbf b_s^H\mathbf R_x\widetilde{\mathbf b}_\theta,
$
$
h_{rr}
=
\widetilde{\mathbf b}_r^H\mathbf R_x\widetilde{\mathbf b}_r,
\qquad
h_{\theta\theta}
=
\widetilde{\mathbf b}_\theta^H\mathbf R_x\widetilde{\mathbf b}_\theta,
\qquad
h_{\theta r}
=
\widetilde{\mathbf b}_\theta^H\mathbf R_x\widetilde{\mathbf b}_r.
$
\end{lemma}
The proof is given in Appendix ~\ref{geometryCRBproof}. 
\begin{figure*}[!b]
\hrulefill
\normalsize
\begin{align}
&\operatorname{CRB}_{r_s}^{(D)}=
\frac{\sigma_r^2}{2|\beta_s|^2TP_t}
\frac{
A_0\big(A_0^2V_\theta+\mu_u\mu_s|\delta_\theta|^2\big)
}{
k^2N^2\Big[
\big(A_0^2V_r+\mu_u\mu_s|\delta_r|^2\big)
\big(A_0^2V_\theta+\mu_u\mu_s|\delta_\theta|^2\big)
-
\big(A_0^2C_{r\theta}+\mu_u\mu_s\Re\{\delta_\theta^*\delta_r\}\big)^2
\Big]
},\nonumber\\
&\operatorname{CRB}_{\theta_s}^{(D)}=
\frac{\sigma_r^2}{2|\beta_s|^2TP_t}
\frac{
A_0\big(A_0^2V_r+\mu_u\mu_s|\delta_r|^2\big)
}{
k^2N^2\Big[
\big(A_0^2V_r+\mu_u\mu_s|\delta_r|^2\big)
\big(A_0^2V_\theta+\mu_u\mu_s|\delta_\theta|^2\big)
-
\big(A_0^2C_{r\theta}+\mu_u\mu_s\Re\{\delta_\theta^*\delta_r\}\big)^2
\Big]
}.
\label{CRbBformula}
\end{align}
\end{figure*}

Note that the centered moments $V_r,V_\theta,C_{r\theta}$ depend only on the target-side
spherical-wave geometry, while the terms
$g_0,g_1,g_2,h_{rr},h_{\theta\theta},h_{\theta r}$ describe how the transmit covariance affects target illumination and how changes in target range and angle are reflected in the received echo. Thus, Lemma \ref{geometryCRBlemma} separates target-side geometry from transmit-covariance effects, showing how each affects the range-angle estimation accuracy. We next specialize the decomposition to a rank-one transmit covariance, which includes the SS beamformer as a special case. In this case, the covariance-dependent terms reduce to a single target-illumination factor.
\begin{lemma}\label{procrbrankone}
For an arbitrary normalized rank-one transmit covariance
$
\mathbf R_x=\mathbf w\mathbf w^H,
$
define
$
g(\phi)\triangleq\mathbf b_s^H(\phi)\mathbf w.
$
Then
$
\operatorname{CRB}_{r_s}
=
\frac{\sigma_r^2}{2|\beta_s|^2TP_t}
\cdot
\frac{V_\theta}
{k^2N|g(\phi)|^2\left(V_rV_\theta-C_{r\theta}^2\right)},
$
and
$
\operatorname{CRB}_{\theta_s}
=
\frac{\sigma_r^2}{2|\beta_s|^2TP_t}
\cdot
\frac{V_r}
{k^2N|g(\phi)|^2\left(V_rV_\theta-C_{r\theta}^2\right)}.
$
\end{lemma}
The proof is given\footnote{To obtain finite range–angle CRBs, we require nonzero $g(\phi)$ and
$
V_r(\phi)>0,
\qquad
V_\theta(\phi)>0,
\qquad
V_r(\phi)V_\theta(\phi)-C_{r\theta}^2(\phi)>0.
$} in Appendix ~\ref{procrbrankoneproof}. For SS, define the target-illumination gain
\begin{align}
g^{(S)}(\phi)
&\triangleq
\mathbf b_s^H(\phi)\mathbf w^{(S)}(c_u,c_s,\phi)
\nonumber\\
&=
\frac{\sqrt N\big(c_s+c_u\rho^*(\phi)\big)}
{\sqrt{|c_u|^2+|c_s|^2+2\Re\{c_u^*c_s\rho(\phi)\}}}.
\end{align}
Substituting $g(\phi)=g^{(S)}(\phi)$ into Lemma~\ref{procrbrankone} yields the corresponding CRBs for SS beamforming.

Lemma~\ref{procrbrankone} shows the key CRB parameters are all
monotonic in $|g(\phi)|^2$ for a fixed target geometry. Hence, at a fixed
rotation, minimizing $\operatorname{CRB}_{r_s}$,
$\operatorname{CRB}_{\theta_s}$,
$\operatorname{tr}(\mathrm{CRB})$, or
$\log\det(\mathrm{CRB})$
reduces to maximizing target illumination subject to the same user quality-of-service (QoS) constraint. The following corollary shows that an optimizer of this constrained
rank-one problem can always be chosen in the span of the user and target
steering vectors.
\begin{corollary}\label{procrbrankoneproofedition}
For a fixed rotation angle $\phi$, consider
$
\max_{\mathbf w}\;|\mathbf b_s^H(\phi)\mathbf w|^2
$
subject to
$
|\mathbf b_u^H(\phi)\mathbf w|^2\ge\Gamma_u$ and $
\|\mathbf w\|^2=1,
$
where $\Gamma_u > 0$ corresponds to a constraint on the user's QoS.
Assume that this optimization problem is feasible.
Then there exists an optimal solution satisfying
$
\mathbf w^\star(\phi)\in
\operatorname{span}\{\mathbf b_s(\phi),\mathbf b_u(\phi)\}.
$
\end{corollary}
The proof is given in Appendix ~\ref{procrbrankoneproofeditionproof}.

Therefore, for every
fixed rotation, the SS family contains an optimal solution of the constrained rank-one
sensing problem. We next derive the corresponding range–angle CRBs for DW.
\begin{lemma}\label{crbBlemma}
For DW, the range and angle CRBs are given by \eqref{CRbBformula}, with
$
A_0=\mu_s+\mu_u|\rho|^2,\qquad
\delta_r=\rho_c-\bar c\rho,$ and $
\delta_\theta=\rho_a-\bar a\rho,
$
where
$
\rho_c=\frac{1}{N}\sum_{n=1}^N c_n b_{u,n}^*b_{s,n}$ and $
\rho_a=\frac{1}{N}\sum_{n=1}^N a_n b_{u,n}^*b_{s,n}.
$
\end{lemma}
The proof is given in Appendix ~\ref{crbBlemmaproof}.

Combining Lemmas~\ref{procrbrankone} and~\ref{crbBlemma} yields the
following common representation of the FIMs for the SS and DW architectures.
\begin{proposition}
\label{prop:derivativediversity}
Let $\mathbf J^{(S)}(\phi)$ and $\mathbf J^{(D)}(\phi)$ denote the FIMs
under the SS and DW architectures, respectively. For SS,
$\mathbf J^{(S)}(\phi)=
\kappa k^2N|g^{(S)}(\phi)|^2\mathbf V_s(\phi)$.
For DW with $0<\mu_u<1$,
$\mathbf J^{(D)}(\phi)=
\kappa k^2N^2
\left[
A_0(\phi)\mathbf V_s(\phi)
+
\frac{\mu_u\mu_s}{A_0(\phi)}
\mathbf D_\delta(\phi)
\right]$.
Moreover, the additional DW term
$\kappa k^2N^2
\frac{\mu_u\mu_s}{A_0(\phi)}
\mathbf D_\delta(\phi)$
is positive semidefinite and is zero if and only if
$\delta_r(\phi)=\delta_\theta(\phi)=0$.
Here, $\kappa\triangleq 2|\beta_s|^2TP_t/\sigma_r^2$,
$\mathbf V_s(\phi)\triangleq
\begin{bmatrix}
V_r(\phi)&C_{r\theta}(\phi)\\
C_{r\theta}(\phi)&V_\theta(\phi)
\end{bmatrix}$,
$\boldsymbol\delta(\phi)\triangleq
[\delta_r(\phi),\delta_\theta(\phi)]^\top$, and
$\mathbf D_\delta(\phi)\triangleq
\Re\!\left\{
\boldsymbol\delta(\phi)\boldsymbol\delta^H(\phi)
\right\}\succeq\mathbf 0$.
\end{proposition}
The proof is given in Appendix ~\ref{prop:derivativediversityproof}. 
This proposition gives the main estimation-side distinction between SS and
DW. SS obtains range--angle information through the scalar target-illumination
factor $|g^{(S)}(\phi)|^2$ together with the target-side moment
$\mathbf V_s(\phi)$. For DW, the term $A_0(\phi)\mathbf V_s(\phi)$ in $\mathbf J^{(D)}(\phi)$ provides
the corresponding illumination--geometry contribution, while the term
$\frac{\mu_u\mu_s}{A_0(\phi)}\mathbf D_\delta(\phi)$ provides
an additional range--angle information.
This term is nonzero
when the communication- and sensing-directed waveforms induce different
changes in the received echo as the target range or angle varies. 
Although this term adds a positive-semidefinite component to the FIM for DW, DW is
not universally superior to SS because the overall CRBs also depend on target
illumination, power allocation, and geometry.
\section{Sensing Performance Analysis: Detection}\label{detectionanalysis}
We next study sensing from the target detection perspective, using detection probability as the performance metric. This complements the estimation perspective considered in Section~\ref{sensingestimation}. Using the same stacked observation vector
$
\mathbf y=\operatorname{vec}(\mathbf Y_s)
$
and signal
$
\mathbf u(\theta_s,r_s)=\operatorname{vec}\!\big(\mathbf G(\phi)\mathbf X\big)
$
defined in Section~\ref{sensingestimation}, the binary hypothesis test for target detection is
$
\mathcal H_0:\mathbf y=\mathbf n,
\qquad
\mathcal H_1:\mathbf y=\beta_s\mathbf u(\theta_s,r_s)+\mathbf n,
\qquad
\mathbf n\sim\mathcal{CN}(\mathbf 0,\sigma_r^2\mathbf I).
$
For a known $\mathbf u(\theta_s,r_s)$ and an unknown complex amplitude $\beta_s$, maximizing the likelihood over $\beta_s$ yields the GLRT statistic
$
T_{\rm GLRT}
=
\frac{|\mathbf u^H\mathbf y|^2}{\sigma_r^2\|\mathbf u\|^2}.
$ Under $\mathcal H_0$,
$
\mathbf u^H\mathbf y
=
\mathbf u^H\mathbf n
\sim
\mathcal{CN}(0,\sigma_r^2\|\mathbf u\|^2),
$
so
$
T_{\rm GLRT}\sim \mathrm{Exp}(1).
$
Hence, for a given threshold $\gamma$, the probability of false alarm is
$
P_{\rm FA}
=
P(T_{\rm GLRT}>\gamma|\mathcal H_0)
=
e^{-\gamma}$ and thus $\gamma=-\ln P_{\rm FA}.
$
Under $\mathcal H_1$,
$
\mathbf u^H\mathbf y
\sim
\mathcal{CN}(\beta_s\|\mathbf u\|^2,\sigma_r^2\|\mathbf u\|^2),
$
so $2T_{\rm GLRT}$ follows a noncentral chi-square distribution with two
degrees of freedom and detection noncentrality parameter
$
\lambda_{\rm det}(\phi)
\triangleq
\frac{2|\beta_s|^2\|\mathbf u\|^2}{\sigma_r^2}.
$
Moreover,
$
\|\mathbf u\|^2
=
\sum_{l=1}^{T}\|\mathbf G(\phi)\mathbf X[l]\|^2
=
TP_t\,T_0(\phi),
$
where
$
T_0(\phi)=\operatorname{tr}\!\big(\mathbf G(\phi)\mathbf R_x\mathbf G^H(\phi)\big).
$
Therefore,
$
\lambda_{\rm det}(\phi)
=
\frac{2|\beta_s|^2TP_t}{\sigma_r^2}T_0(\phi),
$
and the detection probability is
\begin{align}
P_D(\phi)
=
Q_1\!\left(
\sqrt{\frac{2|\beta_s|^2TP_t}{\sigma_r^2}T_0(\phi)},
\sqrt{-2\ln P_{\rm FA}}
\right),
\label{pdexpr}
\end{align}
where $Q_1(\cdot,\cdot)$ is the Marcum $Q$-function. 

For fixed
$P_{\rm FA}$, $P_D(\phi)$ is monotonically increasing in
$\lambda_{\rm det}(\phi)$ and hence in $T_0(\phi)$. Optimizing the rotation angle for detection performance then reduces to maximizing the target-illumination quantity $T_0(\phi)$, which also appears in the FIM in Proposition~\ref{generalCRBprop}. Accordingly, the detection-centric rotation angle for SS is
\begin{align}
\phi_{d}^{(S),\star}(c_u,c_s)
&=
\arg\max_{\phi\in\Phi}T_0^{(S)}(\phi)
=
\arg\max_{\phi\in\Phi}|g^{(S)}(\phi)|^2.
\label{ssdet}
\end{align}
For DW, the detection-centric rotation angle is
\begin{align}
\phi_{d}^{(D),\star}(\mu_u)
&=
\arg\max_{\phi\in\Phi}A_0(\phi)
=
\arg\max_{\phi\in\Phi}|\rho(\phi)|^2,
\label{DWdet}
\end{align}
because
$
T_0^{(D)}(\phi)
=
N^2A_0(\phi)
=
N^2\big(\mu_s+\mu_u|\rho(\phi)|^2\big).
$

The detection-centric rotations in \eqref{ssdet} and \eqref{DWdet} are
obtained for prescribed SS coefficients and DW power fractions. We next
show that, at every rotation, the SS family contains the detection
optimum over all transmit covariances under the same user-QoS constraint.
\begin{lemma}
\label{thm:detection-optimality}
For a fixed rotation angle $\phi$, consider
\begin{align}
\max_{\mathbf R_x\succeq\mathbf 0}\;
P_D(\mathbf R_x,\phi)
\quad\text{s.t.}\quad
\operatorname{tr}(\mathbf R_x)=1,\;\;
\mathbf v_u^H(\phi)\mathbf R_x\mathbf v_u(\phi)\geq q,
\label{eq:det-cov-problem}
\end{align}
where
$P_D(\mathbf R_x,\phi)\triangleq
Q_1\!\big(\sqrt{\tfrac{2|\beta_s|^2TP_t}{\sigma_r^2}T_0(\mathbf R_x,\phi)},
\sqrt{-2\ln P_{\rm FA}}\big)$ is the detection probability at a fixed
$P_{\rm FA}$\footnote{$P_D(\phi)$ in \eqref{pdexpr} is shorthand for
$P_D(\mathbf R_x(\phi),\phi)$.},
and $q\triangleq\gamma_{u,\min}/\tilde\Gamma_u\in(0,1]$ is the normalized
QoS level. Here, $\gamma_{u,\min}$ is the minimum user SINR requirement and
$\tilde\Gamma_u\triangleq P_t|\beta_u|^2N/\sigma_u^2$ is the maximum
achievable user SNR, with $\gamma_{u,\min}\le\tilde\Gamma_u$.
Then \eqref{eq:det-cov-problem} has a rank-one optimal solution
$\mathbf R_x^\star=\mathbf w^\star\mathbf w^{\star H}$ with
$\|\mathbf w^\star\|=1$ and
$\mathbf w^\star\in\operatorname{span}\{\mathbf b_u(\phi),\mathbf b_s(\phi)\}$,
i.e., $\mathbf w^\star$ is an SS beamformer of the form \eqref{beamA}.
\end{lemma}
The proof of Lemma \ref{thm:detection-optimality} is given in Appendix \ref{app:detection-optimality}. Lemma~\ref{thm:detection-optimality} holds pointwise in $\phi$.
Hence it also holds after optimizing over $\phi$, provided that the SS
coefficients $(c_u,c_s)$ are re-optimized at each rotation. It leads to the following
comparison between the two architectures.
\begin{corollary}
\label{cor:detection-dw}
Fix $\phi$ and $\gamma_{u,\min}\le\tilde\Gamma_u$. For any DW power
fraction $\mu_u$ satisfying
$\mathrm{SINR}_u^{(D)}(\mu_u,\phi)\ge\gamma_{u,\min}$, there exist SS
coefficients $(c_u,c_s)$ such that
$\mathrm{SNR}_u^{(S)}(c_u,c_s,\phi)\ge\gamma_{u,\min}$ and
$P_D^{(S)}(c_u,c_s,\phi)\ge P_D^{(D)}(\mu_u,\phi)$.
\end{corollary}
The proof of Corollary \ref{cor:detection-dw} is provided in Appendix \ref{proof:detection-dw}. Thus, under the same user-SINR requirement, the additional sensing waveform in DW does not improve detection: at every fixed rotation, an SS design exists whose detection probability is at least as high as that of any feasible DW design.
\section{Analytical Approximations}\label{approximations}
This
section derives second-order approximations for the rotation-dependent
quantities entering the communication, detection, and estimation metrics.
Specifically, we approximate the correlation quantities
$\rho$, $\rho_a$, and $\rho_c$, together with the target-side
moments $\bar a$, $\bar c$, $V_\theta$, $V_r$, and $C_{r\theta}$. The next two subsections derive approximations for the correlation quantities and the target-side moments, respectively. In the subsequent rotation-design sections, these approximations will be used to derive compact expressions for optimal rotation angles.
\subsection{Communication-Side Quantities}\label{commside}
The user--target steering correlation
$
\rho(\phi)
=
\frac{1}{N}\sum_{n=1}^{N} b_{u,n}^*b_{s,n}
$
enters the communication, detection, and range–angle CRBs. For DW, these CRBs
also depend on the weighted correlations
$
\rho_c(\phi)
=
\frac{1}{N}\sum_{n=1}^{N} c_n b_{u,n}^*b_{s,n}
$
and
$
\rho_a(\phi)
=
\frac{1}{N}\sum_{n=1}^{N} a_n b_{u,n}^*b_{s,n},
$ as defined in Lemma~\ref{crbBlemma}.
Using a second-order phase expansion, the user--target correlation is
approximated as
$
\rho(\phi)
\approx
\frac{1}{N}\sum_{n=1}^{N}
e^{j\alpha(\phi)\chi_n-j\beta(\phi)\chi_n^2},
$
where
\begin{align}
&\alpha(\phi)
=
kd\big[
\cos(\theta_s-\phi)-\cos(\theta_u-\phi)
\big],\nonumber\\
&\beta(\phi)
=
\frac{kd^2}{2}
\left[
\frac{\sin^2(\theta_s-\phi)}{r_s}
-
\frac{\sin^2(\theta_u-\phi)}{r_u}
\right].
\end{align}
Recall that
$
\Delta_s\triangleq\theta_s-\phi.
$
We have
$
a_n
\approx
A\chi_n+B\chi_n^2,
\qquad
c_n
\approx
-C\chi_n^2,
$
where
$
A=d\sin\Delta_s,
\qquad
B=\frac{d^2}{r_s}\sin\Delta_s\cos\Delta_s,
\qquad
C=\frac{d^2}{2r_s^2}\sin^2\Delta_s.
$
Therefore, the weighted correlations are approximated by
\begin{align}
&\rho_a(\phi)
\approx
\frac{1}{N}\sum_{n=1}^{N}
\big(A\chi_n+B\chi_n^2\big)
e^{j\alpha(\phi)\chi_n-j\beta(\phi)\chi_n^2},
\\
&\rho_c(\phi)
\approx
-\frac{C}{N}\sum_{n=1}^{N}
\chi_n^2
e^{j\alpha(\phi)\chi_n-j\beta(\phi)\chi_n^2}.
\end{align}
Thus, $\rho(\phi)$ captures the unweighted user--target coupling, whereas
$\rho_a(\phi)$ and $\rho_c(\phi)$ capture the corresponding
range-- and angle--weighted responses that appear in the CRBs for DW.
When
$\beta(\phi)\approx0$, the unweighted correlation reduces to the Dirichlet form
$
\rho_D(\phi)
\triangleq
\frac{1}{N}\sum_{n=1}^{N}
e^{j\alpha(\phi)\chi_n}
=
\frac{\sin\!\big(N\alpha(\phi)/2\big)}
{N\sin\!\big(\alpha(\phi)/2\big)},
$ with \(\rho_D(\phi)=1\) when \(\alpha(\phi)=0\).
\subsection{Sensing-Side Quantities}\label{sensingside}
The centered target-side moments \(V_\theta\), \(V_r\), and \(C_{r\theta}\) appear in the CRB expressions for both the SS and DW architectures, given in Lemmas~\ref{procrbrankone} and~\ref{crbBlemma}, respectively, while \(\bar a\) and \(\bar c\) enter the CRBs for DW through \(\delta_\theta\) and \(\delta_r\). In the following, we derive second-order approximations for these geometry-related quantities. We assume
$D/r_s\ll1$ and exclude the endfire region
$|\sin(\theta_s-\phi)|\approx0$, where the local range--angle geometry
becomes degenerate.
Using the second-order approximations for $a_n$ and $c_n$ given in Subsection \ref{commside}, we
obtain
$
\bar a
\approx
\frac{d^2(N^2-1)}{12r_s}
\sin\Delta_s\cos\Delta_s,
$
$
\bar c
\approx
-\frac{d^2(N^2-1)}{24r_s^2}
\sin^2\Delta_s,
$
$
C_{r\theta}
\approx
-\frac{d^4(N^2-1)(N^2-4)}
{360r_s^3}
\sin^3\Delta_s\cos\Delta_s,
$
$
V_\theta
\approx
\frac{d^2(N^2-1)}{12}\sin^2\Delta_s
+
\frac{d^4(N^2-1)(N^2-4)}
{180r_s^2}
\sin^2\Delta_s\cos^2\Delta_s,
$
and
$
V_r
\approx
\frac{d^4(N^2-1)(N^2-4)}
{720r_s^4}
\sin^4\Delta_s.
$

Substituting the above moments and correlation approximations into
Lemmas~\ref{geometryCRBlemma}, \ref{procrbrankone}, and
\ref{crbBlemma} gives the corresponding approximate CRBs for an
arbitrary covariance, for rank-one transmission, and for DW, respectively.
For rank-one transmission, Lemma~\ref{procrbrankone} gives
the following second-order approximations for the range and angle CRBs
\begin{align}
\operatorname{CRB}_{r_s}
&\approx
\frac{24\sigma_r^2 r_s^4}
{|\beta_s|^2TP_t k^2Nd^4(N^2-1)
 |g(\phi)|^2\sin^4\Delta_s}
\nonumber\\
&\quad{}\times
\left(
\frac{15}{N^2-4}
+\frac{d^2}{r_s^2}\cos^2\Delta_s
\right),\\
\operatorname{CRB}_{\theta_s}
&\approx
\frac{6\sigma_r^2}
{|\beta_s|^2TP_t k^2Nd^2(N^2-1)
 |g(\phi)|^2\sin^2\Delta_s}.
\end{align}
\section{Communication-Centric Rotation Optimization}
\label{Communicationcentricrotation}
This section analyzes the communication-centric rotation problem separately for the SS and DW architectures. We fix \((c_u,c_s)\) for SS and \(0<\mu_u<1\) for DW, and optimize over \(\phi\). For each architecture, we derive the exact stationary-point condition for the communication-centric rotation optimization, then use the approximation of \(\rho(\phi)\) derived in Subsection~\ref{commside} to obtain approximate expressions for the optimal rotation angles.
\begin{remark}\label{fullcomA}
If, for the SS architecture, the coefficients \((c_u,c_s)\) and the rotation \(\phi\) are optimized for a communication-centric objective, then any unit-norm beamforming vector \(\mathbf w\) satisfies
$\mathrm{SNR}_u(\mathbf w,\phi)
 =
\frac{P_t|\beta_u|^2N}{\sigma_u^2}
\left|\mathbf v_u^H(\phi)\mathbf w\right|^2
\le \frac{P_t|\beta_u|^2N}{\sigma_u^2}.$ Equality is attained for \(\mathbf w=e^{j\psi}\mathbf v_u(\phi)\), which corresponds to \(c_s=0\) and \(c_u\neq0\). Thus, the communication-centric design reduces to the communication-only beamformer, and \(\phi\) can be arbitrary.
Let $\Gamma_0\triangleq P_t|\beta_u|^2N$. For any fixed rotation
$\phi$, the communication SINR for DW is strictly increasing in $\mu_u$,
since
\begin{align}
\frac{\partial}{\partial\mu_u}
\mathrm{SINR}_u^{(D)}(\mu_u,\phi)
=
\frac{
\Gamma_0\big(\Gamma_0|\rho(\phi)|^2+\sigma_u^2\big)
}{
\big((1-\mu_u)\Gamma_0|\rho(\phi)|^2+\sigma_u^2\big)^2
}
>0.
\end{align}
Hence, the communication-centric design selects
$\mu_u^\star=1$ and $\mu_s^\star=0$. In this special case,
$\mathrm{SINR}_u^{(D)}
=
\frac{P_t|\beta_u|^2N}{\sigma_u^2},$
which is independent of $\phi$. Thus, the design reduces to communication-only beamformer, and the rotation angle is arbitrary.
Accordingly, in the rotation analysis that follows, we fix \((c_u,c_s)\) for SS and \(\mu_u\) for DW, with \(0<\mu_u<1\), and optimize only the rotation angle \(\phi\), rather than optimizing the architecture parameters and \(\phi\) at the same time.
\end{remark}
\subsection{Communication-Centric Rotation for SS}
\label{comSS}
The exact
communication-centric rotation for SS is defined by
$
\phi_{u}^{(S),\star}(c_u,c_s)
\in
\arg\max_{\phi\in\Phi}
\mathrm{SNR}_u^{(S)}(c_u,c_s,\phi).
$
Using \eqref{SINRA}, this is equivalent to maximizing
\begin{align}
J^{(S)}(\phi)
\triangleq
\frac{|c_u+c_s\rho(\phi)|^2}
{|c_u|^2+|c_s|^2+2\Re\{c_u^*c_s\rho(\phi)\}}.
\end{align}
Let
$
C_0\triangleq |c_u|^2+|c_s|^2,
\qquad
s_0\triangleq c_u^*c_s.
$
For $c_s\neq0$, any interior stationary point satisfies
\begin{align}
&(1-|\rho(\phi)|^2)\Re\{s_0\rho'(\phi)\}
\nonumber\\
&\quad+
\big(C_0+2\Re\{s_0\rho(\phi)\}\big)
\Re\{\rho^*(\phi)\rho'(\phi)\}
=0.
\label{comsseq}
\end{align}
If $c_s=0$ (and hence $c_u\neq0$), then
$J^{(S)}(\phi)=1$ for all $\phi\in\Phi$. Thus, the communication performance is independent of the rotation angle.
Let
$
\theta_m=\frac{\theta_u+\theta_s}{2},
\qquad
\Delta=\frac{\theta_s-\theta_u}{2}.
$
Using the definitions of $\alpha(\phi)$ and $\rho_D(\phi)$ in
Section~\ref{commside}, we have
$
\alpha(\theta_m)=0
$
and
$
\rho_D(\theta_m)=1.
$
Therefore, 
$
J^{(S)}(\theta_m)
\approx
1,
$
which is the maximum possible value of \(J^{(S)}(\phi)\) for any unit-norm beamformer. Hence, the resulting approximation to the
communication-optimal rotation is
\begin{align}
\phi_{u}^{(S),\mathrm{Dir}}
=\theta_m
=\frac{\theta_u+\theta_s}{2}.
\label{comss_dir}
\end{align}
\subsection{Communication-Centric Rotation for DW}
\label{comDW}
The exact DW
communication-centric rotation is defined by
$
\phi_{u}^{(D),\star}(\mu_u)
\in
\arg\max_{\phi\in\Phi}
\mathrm{SINR}_u^{(D)}(\mu_u,\phi).
$
Since
$
\mathrm{SINR}_u^{(D)}(\mu_u,\phi)
=
\frac{\mu_uP_t|\beta_u|^2N}
{(1-\mu_u)P_t|\beta_u|^2N|\rho(\phi)|^2+\sigma_u^2},
$
the SINR is strictly decreasing in $|\rho(\phi)|^2$. Hence,
$
\phi_{u}^{(D),\star}(\mu_u)
\in
\arg\min_{\phi\in\Phi}|\rho(\phi)|^2,
$ and any interior stationary point satisfies
\begin{align}
\Re\{\rho^*(\phi)\rho'(\phi)\}=0.
\label{commDW_stationary}
\end{align}
Under the approximation $\rho(\phi)\approx\rho_D(\phi)$, the
communication-optimal rotation minimizes $|\rho_D(\phi)|^2$. If $\sin\Delta=0$, then
$\rho_D(\phi)=1$ for all $\phi\in\Phi$, so the communication performance is approximately independent of the rotation angle.
If $\sin\Delta\neq0$, the numerator of $\rho_D(\phi)$,
$\sin\!\big(N\alpha(\phi)/2\big)$, is zero when
$
\alpha(\phi)=\frac{2\pi\ell}{N},\qquad \ell\in\mathbb Z.
$
We exclude $\ell\in N\mathbb Z$, for which the denominator of $\rho_D(\phi)$,
$\sin\!\big(\alpha(\phi)/2\big)$, also vanishes and the ratio is
nonzero. Thus, the zeros of $\rho_D(\phi)$
satisfy
$
\alpha(\phi)=\frac{2\pi\ell}{N},
\qquad
\ell\in\mathbb Z\setminus N\mathbb Z.
$
Since
$
\alpha(\phi)=2kd\sin\Delta\sin(\phi-\theta_m),
$
these angles satisfy
\begin{align}
\sin(\phi-\theta_m)
=
q_\ell,
\qquad
q_\ell
=
\frac{\ell\lambda}{2Nd\sin\Delta},
\label{commdw_dir_condition}
\end{align}
If (\ref{commdw_dir_condition}) has one or
more solutions in $\Phi$, each solution is approximately communication-optimal, since it gives $|\rho_D(\phi)|^2=0$.
With $d=\lambda/2$, we have
$q_\ell=\ell/(N\sin\Delta)$. The smallest nonzero indices in
$\mathbb Z\setminus N\mathbb Z$ are $\ell=\pm1$. A solution to
\eqref{commdw_dir_condition} requires $|q_\ell|\leq1$, so a zero of
$\rho_D(\phi)$ can occur only if $N|\sin\Delta|\geq1$. Conversely,
when $N|\sin\Delta|\geq1$, one of $\ell=\pm1$ gives an angle
$\phi\in\Phi$ satisfying $\sin(\phi-\theta_m)=q_\ell$; hence
$\rho_D(\phi)=0$. Therefore, $\rho_D(\phi)$ has a zero for some
$\phi\in\Phi$ if and only if
\begin{align}
|\theta_s-\theta_u|
&\geq 2\arcsin(1/N).
\label{condition}
\end{align}
If $\sin\Delta\neq0$ and \eqref{condition} fails, the approximate
communication-optimal rotation is obtained by a one-dimensional
numerical minimization of $|\rho_D(\phi)|^2$ over $\Phi$.
\section{Communication--Detection Synergy and Conflict}\label{comdetcompat}
For SS, the communication objective $J^{(S)}(\phi)$ defined in
Subsection~\ref{comSS} and the normalized detection gain
$|g^{(S)}(\phi)|^2/N$ satisfy
\begin{align}
J^{(S)}(\phi)
-
\frac{|g^{(S)}(\phi)|^2}{N}
=
\frac{
(|c_u|^2-|c_s|^2)(1-|\rho(\phi)|^2)
}{
Z^{(S)}(\phi)
}.
\end{align}
Therefore, when $|c_u|=|c_s|$, the communication- and detection-centric objectives are identical
for every rotation angle. Their optimal rotations therefore coincide.
For DW with $0<\mu_u<1$, the communication-centric rotation angle minimizes
$|\rho(\phi)|^2$, as shown in Subsection~\ref{comDW}, whereas the
detection-centric rotation angle maximizes it, as shown in \eqref{DWdet}.
Hence, DW exhibits a direct conflict between communication-centric and sensing-centric optimization.
To quantify the resulting performance loss, define, for any
comparison angle $\phi_j\in\Phi$,
$
L_{d}^{(D)}(\phi_j)
\triangleq
10\log_{10}
\left(
\frac{T_0^{(D)}(\phi_d^{(D),\star})}
{T_0^{(D)}(\phi_j)}
\right)
$
and
$
L_{u}^{(D)}(\phi_j)
\triangleq
10\log_{10}
\left(
\frac{\mathrm{SINR}_u^{(D)}(\phi_u^{(D),\star})}
{\mathrm{SINR}_u^{(D)}(\phi_j)}
\right).
$
Using $0\leq|\rho(\phi)|^2\leq1$, these losses satisfy
\begin{align}
L_{d}^{(D)}
&\leq
10\log_{10}\!\left(\frac{1}{\mu_s}\right),
&
L_{u}^{(D)}
&\leq
10\log_{10}\!\left(1+\mu_s\tilde{\Gamma}_u\right),
\label{compatibility}
\end{align}
where
$
\tilde{\Gamma}_u
=
\frac{P_t|\beta_u|^2N}{\sigma_u^2}.
$
For the balanced power fraction \(\mu_s=0.5\), the detection-information loss is bounded by \(3.01\) dB, whereas the communication-loss bound depends on the communication SNR and can be much larger. Thus, the DW conflict is asymmetric: the communication loss at the detection-centric rotation angle may greatly exceed the detection performance loss at the communication-centric rotation angle.
\section{Sensing-Centric Rotation Optimization}\label{sensingcentric}
This section analyzes the sensing-centric rotation problem separately for the SS and DW architectures. The sensing-centric rotations
$\phi_s^{(S),\star}$ and $\phi_s^{(D),\star}$ are defined in
Section~\ref{problemformulation}. For each architecture, we derive the exact stationary-point condition for the sensing-centric rotation optimization. We then use the second-order approximations for the target-side geometry quantities derived in Subsection~\ref{sensingside} to obtain approximate expressions for the sensing-centric rotation angles and derive their corresponding stationary-point conditions.
\subsection{Sensing-centric Rotation for SS}\label{sensingA}
For SS, substituting $g(\phi)=g^{(S)}(\phi)$ into
Lemma~\ref{procrbrankone} gives
$
\det\!\big(\mathrm{CRB}^{(S)}(\phi)\big)
\propto
\frac{1}{
|g^{(S)}(\phi)|^4
\big(
V_r(\phi)V_\theta(\phi)-C_{r\theta}^2(\phi)
\big)
}.
$
Therefore, the exact sensing-centric rotation angle for SS is equivalently
\begin{align}
\phi_{s}^{(S),\star}(c_u,c_s)
&=\arg\max_{\phi\in\Phi}\Bigl[
2\log |g^{(S)}(\phi)|^2
\nonumber\\
&\qquad+\log\!\left(
V_r(\phi)V_\theta(\phi)-C_{r\theta}^2(\phi)
\right)
\Bigr].
\end{align}
Any interior stationary point satisfies
\begin{align}
2\frac{d}{d\phi}\log |g^{(S)}(\phi)|^2
+
\frac{d}{d\phi}
\log\!\big(
V_r(\phi)V_\theta(\phi)-C_{r\theta}^2(\phi)
\big)
=0.
\label{sensingSSstationary}
\end{align}
Under the second-order approximations for the target-side moments derived in Subsection~\ref{sensingside}, we have
$
V_r(\phi)V_\theta(\phi)-C_{r\theta}^2(\phi)
\approx
K_S\sin^6(\theta_s-\phi),
$
where $K_S>0$ is independent of $\phi$. Hence, the sensing-centric rotation angle for SS is approximately given by
\begin{align}
\phi_{s}^{(S),\mathrm{appr}}(c_u,c_s)
\triangleq
\arg\max_{\phi\in\Phi}
|g^{(S)}(\phi)|^4
\sin^6(\theta_s-\phi).
\label{sssimple}
\end{align}
The corresponding stationary condition is
$
\frac{d}{d\phi}\log |g^{(S)}(\phi)|^2
=
3\cot(\theta_s-\phi).
$
The objective in (\ref{sssimple}) contains two separate contributions: the factor
$|g^{(S)}(\phi)|^4$ describes SS target illumination, while
$\sin^6(\theta_s-\phi)$ describes the target-side range--angle geometry.
If the illumination factor varies slowly with \(\phi\), the geometry term dominates, and the sensing-optimal rotation lies near the target-broadside angle
\begin{align}
\phi_{\rm broad}
\triangleq
\theta_s-\frac{\pi}{2}.
\label{broadside}
\end{align}
Under the same second-order approximation, optimizing \((c_u,c_s)\) and \(\phi\) for the sensing objective yields the sensing-only beamformer \(c_u^\star=0\), \(c_s^\star\neq0\), with \(\phi^\star=\phi_{\rm broad}\).
\begin{remark}
Under communication-only transmission ($c_s=0$), the SS beamformer is steered
along $\mathbf v_u$. The target illumination is then determined by the
user--target steering correlation:
$
|g^{(S)}(\phi)|^2=N|\rho(\phi)|^2.
$
Thus, each sensing CRB contains an illumination factor
$1/(N|\rho(\phi)|^2)$, multiplied by a factor determined by the
target-side geometry.
Under sensing-only transmission (\(c_u=0,\ c_s\neq0\)), the beam is steered along
$\mathbf v_s$, and the user SNR is
$
\mathrm{SNR}_u^{(S)}(0,c_s,\phi)
=
\frac{P_t|\beta_u|^2N}{\sigma_u^2}|\rho(\phi)|^2.
$
Thus, when \(|\rho(\phi)|\) is small, sensing performance under communication-only beamformer, or communication performance under sensing-only beamformer, may be substantially weaker than the performance of the primary function directly served by the beamformer.
\end{remark}
\subsection{Sensing-centric Rotation for DW}\label{senisngB}
For DW, define
$
A_0(\phi)=\mu_s+\mu_u|\rho(\phi)|^2
$
and
\begin{align}
\Delta^{(D)}(\phi)
&=
\bigl(A_0^2V_r+\mu_u\mu_s|\delta_r|^2\bigr)
\times
\bigl(A_0^2V_\theta+\mu_u\mu_s|\delta_\theta|^2\bigr)
\nonumber\\
&\quad-
\bigl(
A_0^2C_{r\theta}
+\mu_u\mu_s\Re\{\delta_\theta^*\delta_r\}
\bigr)^2.
\end{align}
Using Lemma~\ref{crbBlemma}, the exact sensing-centric rotation angle for DW can be
written as
$
\phi_{s}^{(D),\star}(\mu_u)
=
\arg\max_{\phi\in\Phi}
\left[
\log\Delta^{(D)}(\phi)
-
2\log A_0(\phi)
\right].
$
Any interior stationary point
satisfies
\begin{align}
\frac{d}{d\phi}
\left[
\log\Delta^{(D)}(\phi)
-
2\log A_0(\phi)
\right]
=0.
\label{sensingDWstationary}
\end{align}
If the terms involving $\delta_r$ and $\delta_\theta$
in $\Delta^{(D)}(\phi)$ are small compared with the corresponding
$A_0^2$-weighted geometry terms, then
$
\Delta^{(D)}(\phi)
\approx
A_0(\phi)^4
\big(
V_r(\phi)V_\theta(\phi)-C_{r\theta}^2(\phi)
\big).
$
Under the second-order approximations for the target-side moments derived in Subsection~\ref{sensingside}, the sensing-centric rotation angle for DW is approximately
given by
\begin{align}
\phi_{s}^{(D),\mathrm{appr}}(\mu_u)
\triangleq
\arg\max_{\phi\in\Phi}
A_0(\phi)^2\sin^6(\theta_s-\phi).
\label{DWsimple}
\end{align}
The corresponding stationary condition is
$
\frac{d}{d\phi}\log A_0(\phi)
=
3\cot(\theta_s-\phi),
$
or equivalently
\begin{align}
\frac{
\mu_u\frac{d}{d\phi}|\rho(\phi)|^2
}{
\mu_s+\mu_u|\rho(\phi)|^2
}
=
3\cot(\theta_s-\phi).
\end{align}
The objective in (\ref{DWsimple}) contains two separate contributions: the factor
$A_0(\phi)^2$ describes DW target illumination, while
$\sin^6(\theta_s-\phi)$ describes the target-side range--angle geometry.
If $A_0(\phi)$ varies slowly with \(\phi\), the geometry term dominates, and the sensing-centric rotation is the target-broadside angle
$\phi_{\rm broad}$ defined in \eqref{broadside}.
\section{Special Cases and Explicit Optimal Rotation}\label{specialcases}
In general, the exact communication-centric rotation problems (\eqref{comsseq} and \eqref{commDW_stationary}) and sensing-centric rotation problems (\eqref{sensingSSstationary} and \eqref{sensingDWstationary}) each require a one-dimensional optimization over \(\phi\).
In this section, we first derive exact
closed-form optimal rotation angles. When \(r_u=r_s\), the communication-centric rotation angle for SS and detection-centric rotation angle for DW are both \((\theta_u+\theta_s)/2\). When \(\theta_u=\theta_s=\theta_0\) and \(r_u,r_s>(N-1)d/2\), both rotations equal \(\theta_0\). We then show that, under the second-order approximation, target broadside is sensing-optimal for sensing-only transmission and for a co-located user and target.
\begin{corollary}\label{cor:exactoverlap}
1) Assume
$
r_u=r_s
$.
Then the communication-centric optimal rotation angle for SS is given by
$
\phi_{u}^{(S),\star}(c_u,c_s)= \frac{\theta_u+\theta_s}{2}.
$

2) Assume
$
\theta_u=\theta_s=\theta_0,
\qquad
r_u,r_s>\frac{N-1}{2}d.
$
Then the communication-centric optimal rotation angle for SS is given by
$
\phi_{u}^{(S),\star}(c_u,c_s)= \theta_0.
$

3) For DW,
the detection-centric optimal rotation angle is given by:

(a) $\phi_{d}^{(D),\star}(\mu_u)
=\frac{\theta_u+\theta_s}{2}$, if $r_u=r_s$.

(b) $\phi_{d}^{(D),\star}(\mu_u)
=\theta_0$, if $\theta_u=\theta_s=\theta_0$ and
$r_u,r_s>\frac{N-1}{2}d$.
\end{corollary}
The proof is given\footnote{Cases 2 and 3(b) do not claim range--angle
estimation optimality. At $\phi=\theta_0$, the joint range--angle FIM
is singular, so no finite range--angle CRB is obtained.} in Appendix~\ref{app:exactoverlap}.
The four cases reduce to two geometries, each giving the user and target
the same spatial signature at the specified rotation: Cases 1 and 3(a)
have equal ranges and use the midpoint rotation, while Cases 2 and 3(b)
have a common direction and use the rotation aligned with that
direction. Thus, in Cases 1 and 2, the SS beam aligns with the user channel
and achieves its maximum communication gain. In Cases 3(a) and 3(b), the
same signature maximizes DW target illumination and detection. When
$\mu_s>0$, it also maximizes the sensing signal's gain at the user.
Reducing that interference requires a rotation with smaller $|\rho|$,
which lowers DW detection performance from its maximum.
\begin{corollary}\label{cor:secondorderest}
Under the second-order approximation in Section~\ref{sensingside}.

1) In sensing-only transmission, i.e., $
c_u=0,\qquad c_s\neq0,
$ for SS and $\mu_u=0$ for DW, the
target-broadside rotation minimizes both the range and angle CRBs for
both architectures:
$
\phi_{s}^{(S),\mathrm{appr}}(0,c_s)
=
\phi_{s}^{(D),\mathrm{appr}}(0)
=
\theta_s-\frac{\pi}{2}
$.

2) Assume
$
(\theta_u,r_u)=(\theta_s,r_s).
$
Then, for both SS and DW, we have
$
\mathbf b_u(\phi)=\mathbf b_s(\phi)
\quad\text{and}\quad
\rho(\phi)=1,
\qquad \forall\phi\in\Phi.
$
Hence the communication and detection performance metrics are independent of $\phi$ and the range and angle CRBs are
minimized at target broadside. Thus, the target-broadside rotation is
sensing-optimal without changing communication or detection performance.
\end{corollary}
The proof is given in Appendix~\ref{app:secondorderest}. Under the second-order approximation,
$V_rV_\theta-C_{r\theta}^2$ is proportional to
$\sin^6(\theta_s-\phi)$, and this factor is largest at
$\phi=\theta_s-\pi/2$. In the co-located case,
$\rho(\phi)=1$ for every rotation, so changing $\phi$ does not change
the user-target steering correlation or the communication and detection
metrics. The rotation is therefore selected by the target-side
range--angle geometry, which favors broadside in the cases considered.
\begin{remark}
The broadside optimality result above is consistent with \cite{Zha26}, which shows that target-broadside is optimal for isotropic and sensing-only transmission under the corresponding second-order analysis. Here it appears as a special case of the present framework.
\end{remark}
\section{Robustness to Target-State Perturbation}\label{robustness}
This section extends the preceding rotation analysis by considering uncertainty in the transmitter’s target-state estimate.
The transmitter designs the rotation angle and transmit waveform using an estimated target state, whereas the true target state is modeled as
a random perturbation around that estimate.
We first show how this uncertainty affects communication, and then we derive
robust sensing-centric and detection-centric rotation angles.

Let
$
\hat{\boldsymbol\psi}_s
=
[\hat r_s,\hat\theta_s]^\top
$
denote the estimated target state used for design and let
$
\boldsymbol\psi_s
=
[r_s,\theta_s]^\top
=
\hat{\boldsymbol\psi}_s+\Delta\boldsymbol\psi
$
denote the true state, where
$
\Delta\boldsymbol\psi
=
[\Delta r_s,\Delta\theta_s]^\top.
$
We assume
$
\Delta\boldsymbol\psi
\sim
\mathcal N(\mathbf 0,\mathbf\Sigma_\psi),
$
with
$
\mathbf\Sigma_\psi
=
\begin{bmatrix}
\sigma_{\Delta r}^2 & \sigma_{r\theta}\\
\sigma_{r\theta} & \sigma_{\Delta\theta}^2
\end{bmatrix}.
$
In the following, for notational simplicity we will write $\Delta$ in place of $\Delta\boldsymbol\psi$. We model $\beta_s$ as an unknown complex scalar that remains fixed
across target-state perturbations. Thus, $\Delta$ changes the
element-dependent steering response but not the common round-trip gain.
Define
$
\hat{\mathbf v}_s(\phi)
=
\frac{\mathbf b(\hat\theta_s,\hat r_s,\phi)}{\sqrt N},
\qquad
\mathbf v_s(\phi,\Delta)
=
\frac{\mathbf b(\hat\theta_s+\Delta\theta_s,
\hat r_s+\Delta r_s,\phi)}{\sqrt N},
$
and
$
\hat\rho(\phi)=\mathbf v_u^H(\phi)\hat{\mathbf v}_s(\phi),
\qquad
\rho_\Delta(\phi)
=
\mathbf v_u^H(\phi)\mathbf v_s(\phi,\Delta),
\qquad
\eta(\phi,\Delta)
=
\hat{\mathbf v}_s^H(\phi)\mathbf v_s(\phi,\Delta).
$
The SS and DW designs based on the estimated target state use $\hat{\mathbf v}_s$ in place of
$\mathbf v_s$. In particular,
$
\hat Z^{(S)}
=
|c_u|^2+|c_s|^2
+2\Re\{c_u^*c_s\hat\rho(\phi)\},
$
and
$
\mathbf R_x^{(D)}(\phi)
=
\mu_u\mathbf v_u\mathbf v_u^H
+
\mu_s\hat{\mathbf v}_s\hat{\mathbf v}_s^H.
$ Because the communication metric depends on the design based on the estimated target state
and not on the target-state perturbation $\Delta$, its
rotation optimization retains the nominal form, i.e., the solutions correspond to those in
Sections~\ref{comSS} and~\ref{comDW}, but where the target-dependent terms are evaluated at $(\hat r_s,\hat\theta_s)$ (equivalently, replacing $\rho$ with
$\hat\rho$ and $(r_s,\theta_s)$ with $(\hat r_s,\hat\theta_s)$).
\subsection{Robust Sensing-centric Rotation angle}
For each realization of $\Delta$ and rotation $\phi$, we
evaluate the CRB at the true target state
$(\hat r_s+\Delta r_s,\hat\theta_s+\Delta\theta_s)$, using a transmit
design based on the nominal state, i.e., the estimated target state
$(\hat r_s,\hat\theta_s)$. For each architecture $A\in\{S,D\}$, define the robust
sensing-centric rotation as
$
\bar\phi_s^{(A),\star}
=
\arg\max_{\phi\in\Phi}
\mathbb E_\Delta
\left[
-\log\det\!\big(\operatorname{CRB}^{(A)}(\phi,\Delta)\big)
\right].
$ The following lemma gives second-order approximations to the robust sensing-centric rotation angles.
\begin{lemma}
\label{robustsensinglemma}
Under the second-order approximation, the robust
sensing-centric rotation angles are approximated by
$
\bar\phi_s^{(S),\mathrm{appr}}
=
\arg\max_{\phi\in\Phi}
\{
\mathbb E_\Delta
\left[
\log|g^{(S)}(\phi,\Delta)|^4
\right]
+
\mathcal G_6(\hat\theta_s-\phi)
\}$ and $
\bar\phi_s^{(D),\mathrm{appr}}
=
\arg\max_{\phi\in\Phi}
\{
\mathbb E_\Delta
\left[
\log A_0(\phi,\Delta)^2
\right]
+
\mathcal G_6(\hat\theta_s-\phi)
\},$ 
where
$
g^{(S)}(\phi,\Delta)
\triangleq
\frac{\sqrt N\big(
c_u\rho_\Delta^*(\phi)+c_s\eta^*(\phi,\Delta)
\big)}
{\sqrt{\hat Z^{(S)}}}, 
A_0(\phi,\Delta)
\triangleq
\mu_u|\rho_\Delta(\phi)|^2+
\mu_s|\eta(\phi,\Delta)|^2,$ and $
\mathcal G_6(x)
\triangleq
\mathbb E_{\Delta\theta_s}
\left[
6\log\left|\sin(x+\Delta\theta_s)\right|
\right],
$ respectively.
\end{lemma}
The proof is given in Appendix~\ref{app:robustsensing}.
\subsection{Robust Detection-Centric Rotation angle}
Under target-state uncertainty, the detector performs a GLRT using the estimated target state $(\hat r_s,\hat\theta_s)$. For a
perturbation $\Delta$, the true target state is
$(\hat r_s+\Delta r_s,\hat\theta_s+\Delta\theta_s)$. Define
$
\hat{\mathbf G}(\phi)
=
\hat{\mathbf b}_s(\phi)\hat{\mathbf b}_s^H(\phi)$ and $
\mathbf G_\Delta(\phi)
=
\mathbf b_s(\phi,\Delta)\mathbf b_s^H(\phi,\Delta),
$
where $\mathbf b_s(\phi,\Delta)$ is the steering vector at
$(\hat r_s+\Delta r_s,\hat\theta_s+\Delta\theta_s)$. The vector that is used in GLRT is
$\hat{\mathbf u}(\phi)=\operatorname{vec}(\hat{\mathbf G}(\phi)\mathbf X)$,
whereas the received signal is
$\mathbf y=\beta_s\mathbf u_\Delta(\phi)+\mathbf z$, with
$\mathbf u_\Delta(\phi)=\operatorname{vec}(\mathbf G_\Delta(\phi)\mathbf X)$.
The resulting noncentrality parameter, conditional on 
$\Delta$, is
\begin{align}
\lambda_{\rm det}^{\rm mis}(\phi,\Delta)
&=
\frac{2|\beta_s|^2TP_t}{\sigma_r^2}
\frac{
\left|
\operatorname{tr}\!\left(
\hat{\mathbf G}(\phi)\mathbf G_\Delta(\phi)\mathbf R_x(\phi)
\right)
\right|^2
}{
\operatorname{tr}\!\left(
\hat{\mathbf G}(\phi)\mathbf R_x(\phi)
\hat{\mathbf G}(\phi)
\right)
}.
\label{robustmisnoncentrality}
\end{align}
Thus,
$
P_D(\phi\mid\Delta)
=
Q_1\!\left(
\sqrt{\lambda_{\rm det}^{\rm mis}(\phi,\Delta)},
\sqrt{-2\ln P_{\rm FA}}
\right).
$
For $A\in\{S,D\}$, define the exact robust detection-centric rotation by
$
\bar\phi_d^{(A),\star}
=
\arg\max_{\phi\in\Phi}
\mathbb E_\Delta
\left[
P_D^{(A)}(\phi\mid\Delta)
\right]
$
\footnote{In the numerical results, this expectation is evaluated using
Monte Carlo samples.}. The following lemma gives a low-complexity
approximation for the robust detection-centric rotation angles. 
\begin{lemma}
\label{robustdetectionlemma}
The robust detection-centric
rotation angles are approximated by
$
\bar\phi_d^{(A),\mathrm{appr}}
=
\arg\max_{\phi\in\Phi_{\rm valid}}
\mathcal I_{\rm det}^{(A)}(\phi)
\bigl(1-\upsilon(\phi)\bigr),
\qquad A\in\{S,D\},
$
where $
\Phi_{\rm valid}
\triangleq
\{\phi\in\Phi:0\leq\upsilon(\phi)<1\}
$, $\upsilon(\phi)
\triangleq
k^2\Bigl(
\hat V_r(\phi)\sigma_{\Delta r}^2
+\hat V_\theta(\phi)\sigma_{\Delta\theta}^2
+2\hat C_{r\theta}(\phi)\sigma_{r\theta}
\Bigr),$ $\mathcal I_{\rm det}^{(S)}(\phi)
\triangleq |\hat g^{(S)}(\phi)|^2,
\qquad
\mathcal I_{\rm det}^{(D)}(\phi)
\triangleq \mu_s+\mu_u|\hat\rho(\phi)|^2$, and $\hat g^{(S)}(\phi)
\triangleq
\frac{
\sqrt N\bigl(c_u\hat\rho^*(\phi)+c_s\bigr)
}{
\sqrt{\hat Z^{(S)}(c_u,c_s,\phi)}
}$.
The hatted geometric quantities are evaluated at
$(\hat r_s,\hat\theta_s)$.
\end{lemma}
The proof is given in Appendix~\ref{app:robustdetection}.
Here, $\mathcal I_{\rm det}^{(A)}(\phi)$ is the target-illumination
factor at the nominal state. \(\upsilon(\phi)\) approximates the loss in steering-vector alignment caused by target-state errors. Thus, \(1-\upsilon(\phi)\) approximates the alignment retained between the steering vectors at the estimated and true target states. The product $I_{\rm det}^{(A)}(\phi)
\big(1-\upsilon(\phi)\big)$ favors rotations that provide strong
illumination at the estimated state and are less sensitive to the
state error. When $\mathbf\Sigma_\psi=\mathbf0$, the target-state error is zero and
$\upsilon(\phi)=0$. Thus, the objective being maximized reduces to
$\mathcal I_{\rm det}^{(A)}(\phi)$, evaluated at
$(\hat r_s,\hat\theta_s)$. Hence, its maximizing rotation is the
detection-centric rotation from the preceding analysis, evaluated at
$(r_s,\theta_s)=(\hat r_s,\hat\theta_s)$.
\section{Numerical Results}\label{numerical}
Unless otherwise stated, we adopt $f_c=28$ GHz, $N=129$, $d=\lambda/2$
(here $\lambda\approx10.71$ mm and $D\approx0.686$ m), $r_u=15$ m,
$r_s=20$ m, $c_u=c_s=1$ for SS, and
$\mu_u=\mu_s=0.5$ for DW. Both ranges lie within the radiative
near-field interval. We set $P_t=1$, $\sigma_u^2=10^{-3}$,
$\sigma_r^2=1$, and normalized channel gains $|\beta_u|^2=|\beta_s|^2=1$, and $T=30$. All numerical results use exact finite-array spherical-wave responses, with element-wise path-loss variation neglected. The fixed-orientation baseline is $\phi=0$, and
$
\operatorname{RCRB}_{r_s}=\sqrt{\operatorname{CRB}_{r_s}},
\qquad
\operatorname{RCRB}_{\theta_s}
=\sqrt{\operatorname{CRB}_{\theta_s}}.
$ For the results regarding robust design, we set $\theta_u=40^\circ$, $\hat\theta_s=45^\circ$,
$\hat r_s=20$ m, $\sigma_{\Delta r}=2\tau$ m,
$\sigma_{\Delta\theta}=8\tau^\circ$, and
$\sigma_{r\theta}=0.5\sigma_{\Delta r}\sigma_{\Delta\theta}$.
\begin{figure}[!t]
 \centering
\subfigure[$\operatorname{RCRB}_{r_s}$ versus $\theta_s$.]{\includegraphics[scale=.21]{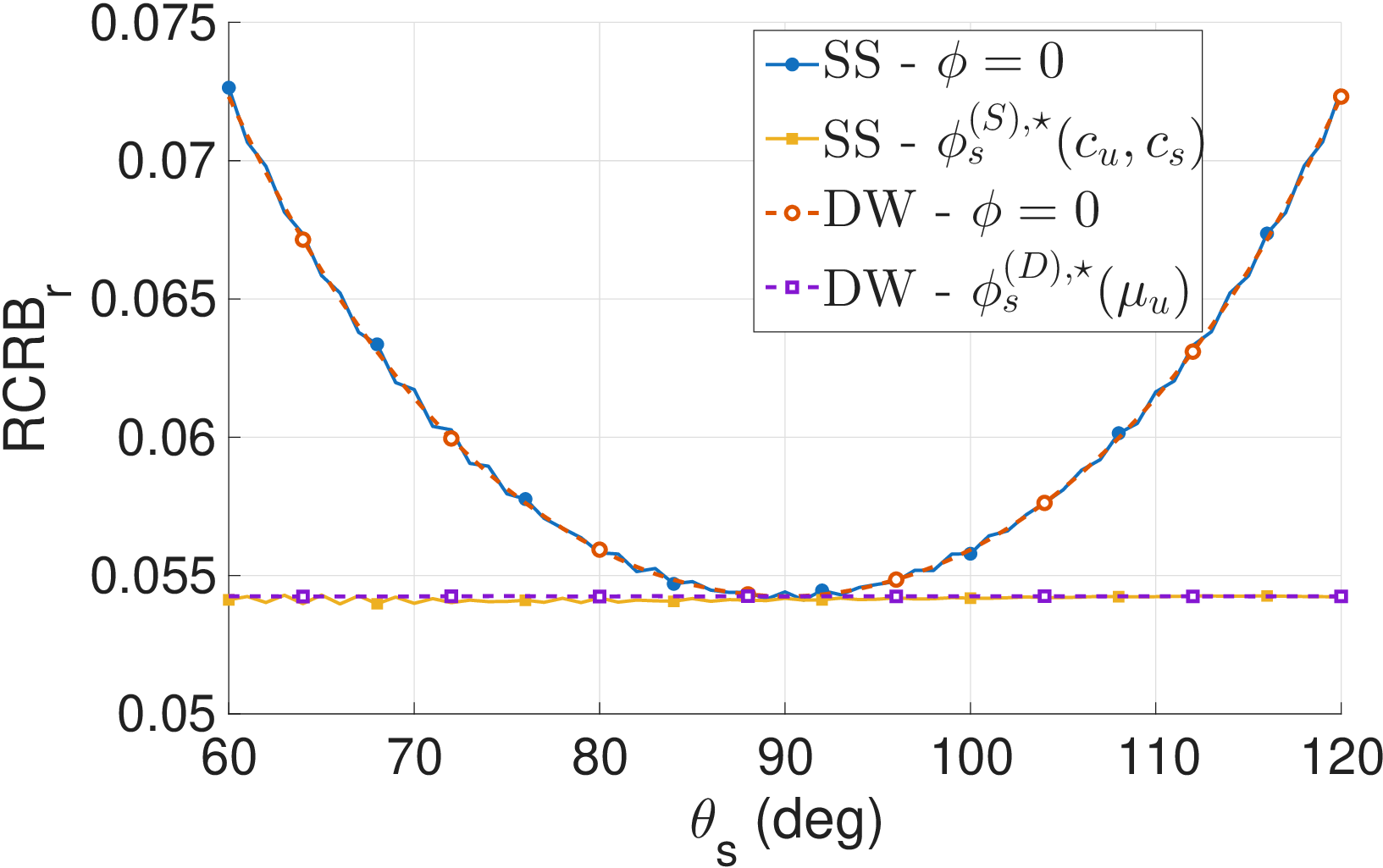}}
\subfigure[$\operatorname{RCRB}_{r_s}$ versus $r_s$.]{\includegraphics[scale=.21]{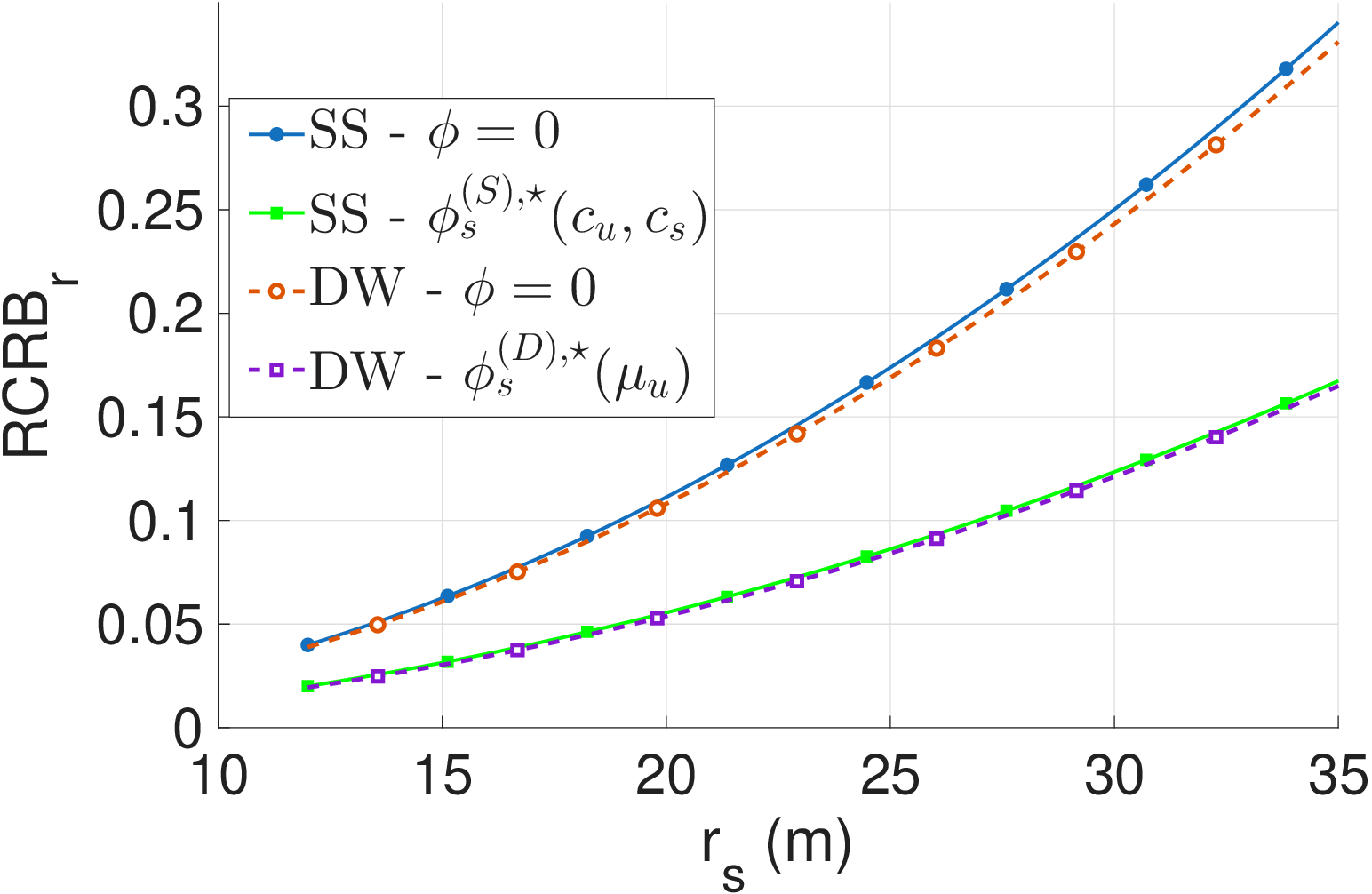}}
\caption{SS and DW: Range RCRB versus $\theta_s$ and $r_s$}
  \label{sweep}
\end{figure}
\begin{figure}[!t]
 \centering
    \includegraphics[scale=.22]{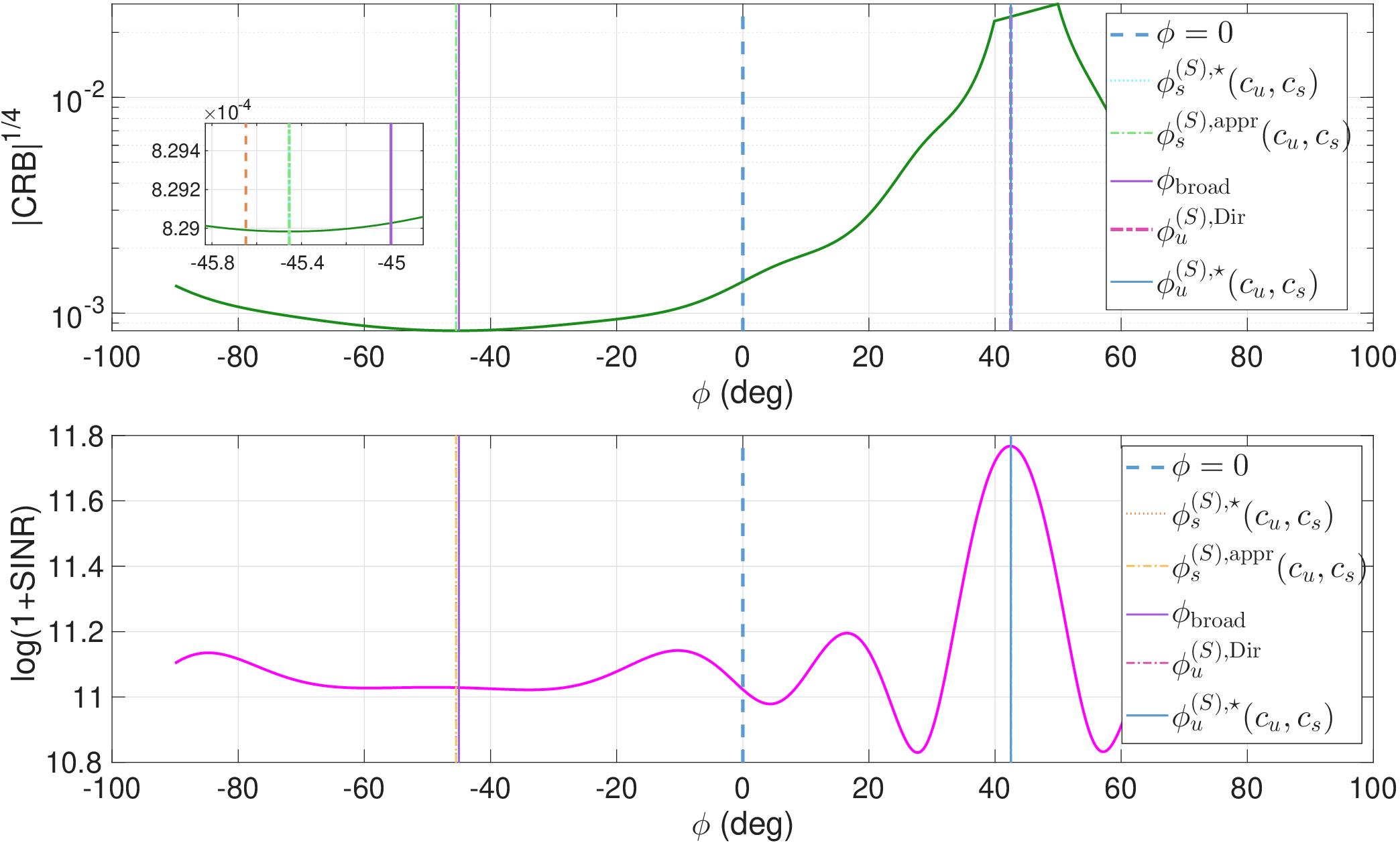}
   \caption{SS estimation and communication performance versus the array
rotation angle $\phi$. The upper plot shows the CRB-based sensing metric
$\det(\operatorname{CRB}(\phi))^{0.25}$, while the lower plot shows the
achievable rate $\log(1+\mathrm{SNR}_u^{(S)}(\phi))$.}
   \label{versusphi}
\end{figure}

Fig.~\ref{sweep}(a) and Fig.~\ref{sweep}(b) show the range RCRB versus
$\theta_s$ and $r_s$, respectively, for both SS and DW. In both figures
and for both architectures, allowing array rotation provides a clear
sensing gain compared with the fixed-orientation baseline $\phi=0$.
In Fig.~\ref{sweep}(b), the exact sensing-centric rotations
$\phi_s^{(S),\star}(c_u,c_s)$ and
$\phi_s^{(D),\star}(\mu_u)$ are characterized by the stationary
conditions \eqref{sensingSSstationary} and
\eqref{sensingDWstationary}, respectively. Comparing the corresponding
exact CRB objectives at these candidates reduces
$\operatorname{RCRB}_{r_s}$ by approximately one half relative to
$\phi=0$ for both architectures. The corresponding reduction in
$\operatorname{RCRB}_{\theta_s}$ is more moderate
\footnote{Since the angular improvement is consistently smaller than the
range improvement, the corresponding angle-RCRB subfigures are omitted for
brevity.}, approximately $29\%$--$31\%$. By contrast,
Fig.~\ref{sweep}(a) shows that rotation gains are small near target
broadside but substantial for off-broadside targets.

The upper and lower subfigures of Fig.~\ref{versusphi} show the
SS CRB-based sensing metric
$\det(\operatorname{CRB}(\phi))^{0.25}$
and the SS achievable rate
$\log(1+\operatorname{SNR}_u^{(S)}(\phi))$,
respectively, versus rotation angle $\phi$. The vertical markers indicate
the rotation angles obtained from the different design criteria. The
exact sensing rotation $\phi_s^{(S),\star}(c_u,c_s)$, the approximate
sensing rotation angle $\phi_s^{(S),\mathrm{appr}}(c_u,c_s)$ given by
\eqref{sssimple}, and the target-broadside rotation
$\phi_{\rm broad}$ given by \eqref{broadside} are very close to one
another. Similarly, the exact communication rotation
$\phi_u^{(S),\star}(c_u,c_s)$ is characterized by \eqref{comsseq}, whereas
the approximation $\phi_u^{(S),\mathrm{Dir}}$ is given by
\eqref{comss_dir}. The close agreement demonstrates the accuracy of the
proposed approximations.
\footnote{The DW results show similar behavior. Due to space limitations,
only the results for SS beamforming are shown.}
The sensing metric is minimized near $\phi_{\rm broad}$, whereas the
communication performance is maximized near the constructive
communication rotation, which is close to
$\phi_u^{(S),\mathrm{Dir}}$ in the considered geometry. However, the
communication curve is relatively flat around the sensing-favorable
region. Consequently, using
$\phi_s^{(S),\star}(c_u,c_s)$ causes only a modest communication penalty.
\begin{figure}[!t]
 \centering
    \includegraphics[scale=.2]{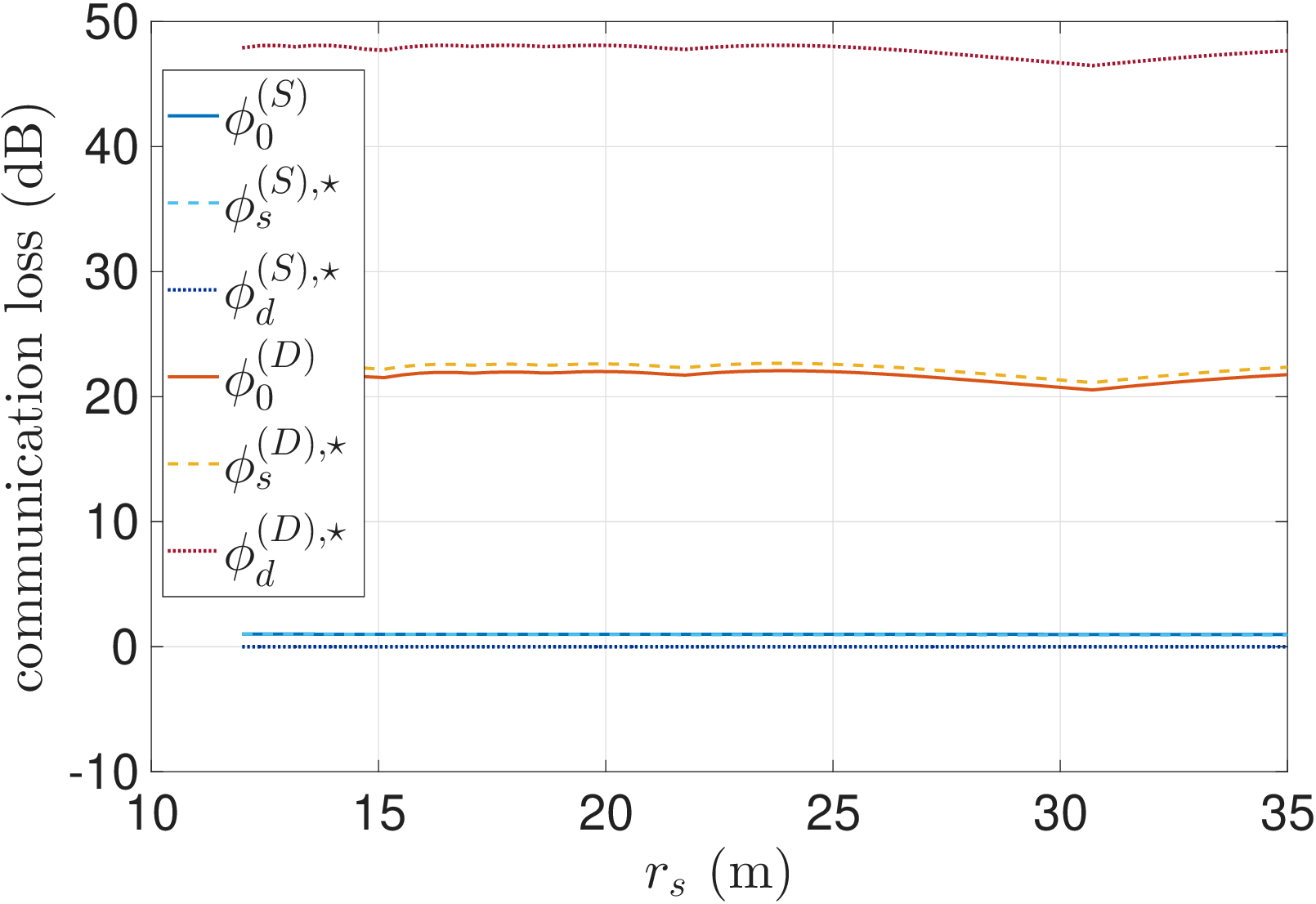}
  \caption{SS and DW: communication loss versus $r_s$.}
   \label{commloss}
\end{figure}
\begin{figure}[!t]
 \centering
    \includegraphics[scale=.26]{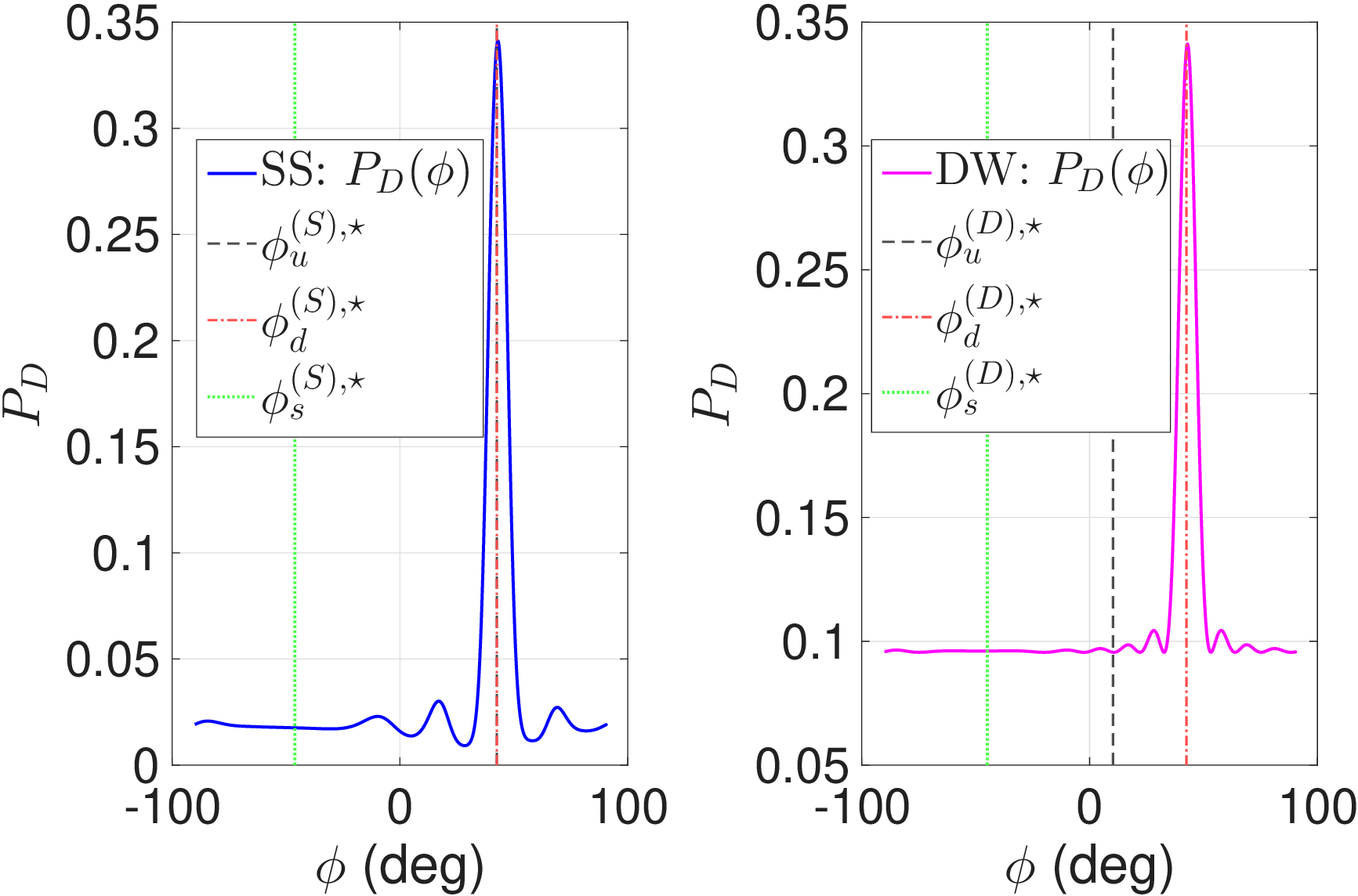}
\caption{Detection probability $P_D$ versus the array rotation angle
$\phi$. The
left and right subfigures correspond to SS and DW, respectively. The
parameters are $P_{\rm FA}=10^{-3}$. For DW, $\mu_u=\mu_s=0.5$; for SS,
$c_u=1$ and \(c_s=0.5e^{\mathrm{j}\pi/4}\).}
   \label{fig:detection}
\end{figure}

Figures~\ref{commloss} and~\ref{fig:detection} compare the rotation
preferences under common geometry and SS/DW parameters:
$c_u=1$, \(c_s=0.5e^{\mathrm{j}\pi/4}\) for SS, and
$\mu_u=\mu_s=0.5$ for DW. Fig.~\ref{commloss} shows communication
loss versus $r_s$ at $P_t=1$, while Fig.~\ref{fig:detection} shows detection
probability versus $\phi$ at $r_s=20~\mathrm{m}$ and
$P_{\rm FA}=10^{-3}$. To avoid saturation, the detection-probability
curves use the reduced power $P_t^{\rm det}=10^{-5}$. This
changes the $P_D$ values, but not the detection-optimal
rotation or the detection-information losses defined as $
L_d^{(A)}(\phi_j)
\triangleq
10\log_{10}\!\left(
\frac{T_0^{(A)}(\phi_d^{(A),\star})}
{T_0^{(A)}(\phi_j)}
\right),
$ $A\in\{S,D\}$.
For any comparison rotation $\phi_j\in\Phi$, define
the communication loss
as
$
L_u^{(A)}(\phi_j)
\triangleq
10\log_{10}\!\left(
\frac{\mathrm{SINR}_u^{(A)}(\phi_u^{(A),\star})}
{\mathrm{SINR}_u^{(A)}(\phi_j)}
\right)
$. The communication-centric rotations for SS and DW, $\phi_u^{(S),\star}(c_u,c_s)$ and
$\phi_u^{(D),\star}(\mu_u)$, are characterized by
\eqref{comsseq} and \eqref{commDW_stationary}, respectively.
For SS, the communication- and detection-centric rotations
$\phi_u^{(S),\star}(c_u,c_s)$ and
$\phi_d^{(S),\star}(c_u,c_s)$ coincide near
$42.50^\circ$. Consequently, the detection-centric
rotation produces essentially zero communication loss and also maximizes
the detection probability. In contrast, the sensing-centric rotation
is near $-46.14^\circ$. This rotation causes a
communication loss of approximately $0.98$ dB and a
detection-information loss of approximately $7.12$ dB. Thus, for the given geometry, communication and detection for SS remain compatible
even with unequal-magnitude coefficients, whereas estimation favors a
substantially different rotation but remains relatively close in
communication performance. For DW, the communication-centric, sensing-centric $\phi_s^{(D),\star}(\mu_u)$ (characterized by \eqref{sensingDWstationary}), and
detection-centric rotations $\phi_d^{(D),\star}(\mu_u)$ (characterized by \eqref{DWdet}) are approximately $10.27^\circ$,
$-44.90^\circ$ (generally close to the
target-broadside rotation $\phi_{\rm broad}$), and $42.50^\circ$, respectively. Using the sensing-centric rotation causes a communication loss of
approximately $22.6$ dB, while using the detection-centric rotation
causes a much larger loss of approximately $48.1$ dB. Moreover,
Fig.~\ref{fig:detection} shows that the detection-information losses at the communication- and
sensing-centric rotations are approximately $3$ dB in this case. 
These values are consistent with the bound in
\eqref{compatibility}.
The results show the exact communication--detection preference conflict for DW and the additional geometry-dependent estimation tradeoff.
\begin{figure}[!t]
 \centering
    \includegraphics[scale=.3]{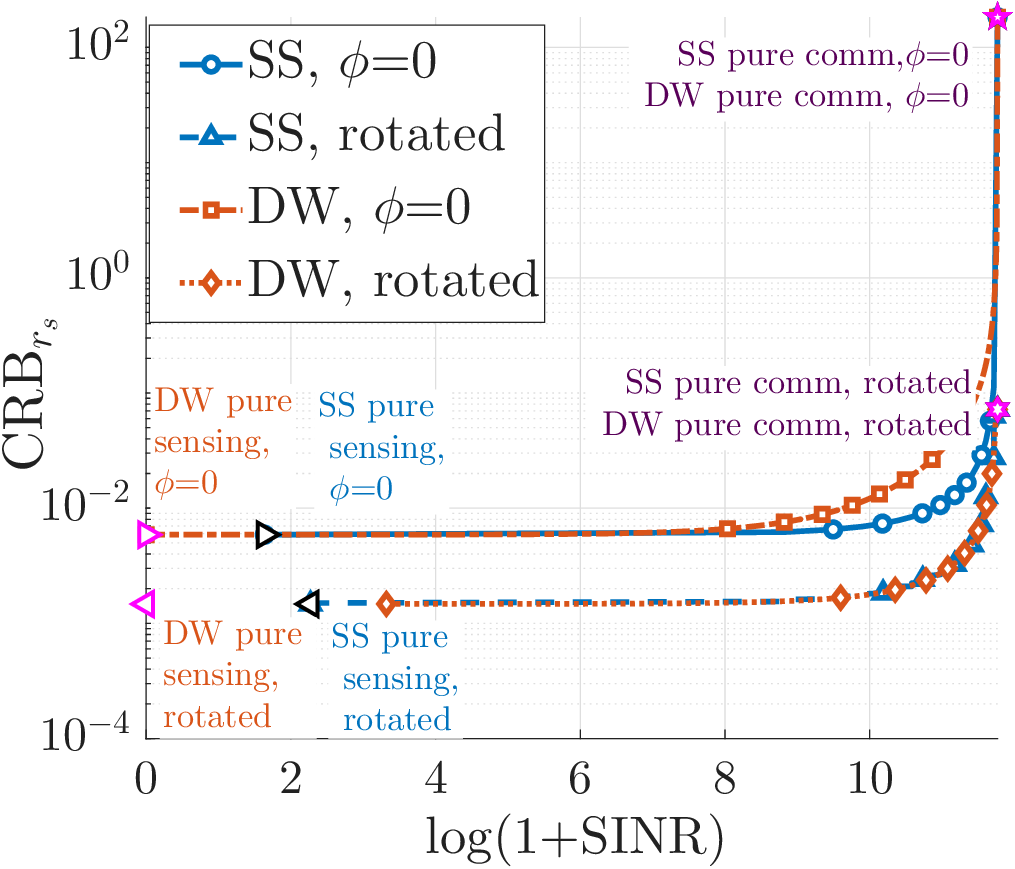}
  \caption{Achievable rate - range CRB tradeoffs for
SS and DW architectures with
$\theta_u=0^\circ$ and $\theta_s=45^\circ$.}
   \label{tradoff}
\end{figure}

Fig.~\ref{tradoff} shows the achievable rate - range CRB tradeoffs, with $\log(1+\mathrm{SINR})$ on the horizontal axis and
$\mathrm{CRB}_{r_s}$ on the vertical axis; hence, better operating points
lie toward the lower-right corner. The boundaries are generated by varying
the rotation and architecture parameters. For SS, we use
$c_u=\sqrt{\xi}$ and
$c_s=\sqrt{1-\xi}e^{\mathrm{j}\psi}$, with $0\leq\xi\leq1$, whereas DW
uses $\mu_u$ with $\mu_s=1-\mu_u$.
Allowing rotation shifts the achievable boundaries substantially downward
relative to the fixed orientation $\phi=0$, while preserving the maximum
achievable rate. At the communication-only endpoint, rotation reduces
$\mathrm{CRB}_{r_s}$ from $1.83\times10^{2}$ to $7.20\times10^{-2}$~m$^2$,
and at the sensing-only endpoint it provides a reduction of approximately
$75\%$. Both designs share the same sensing-only covariance
$\mathbf v_s\mathbf v_s^H$ and hence the same sensing-only range CRB. DW
attains this CRB at zero rate, whereas SS attains the same CRB at a positive rate at that same rotation. Thus, the
zero-rate DW point is dominated by the corresponding SS point. Rotation
therefore improves range estimation across the entire tradeoff without
reducing the communication ceiling. The corresponding angle-CRB curves
show the same qualitative trend and are omitted for brevity.
\begin{figure}[!t]
 \centering
    \includegraphics[scale=.28]{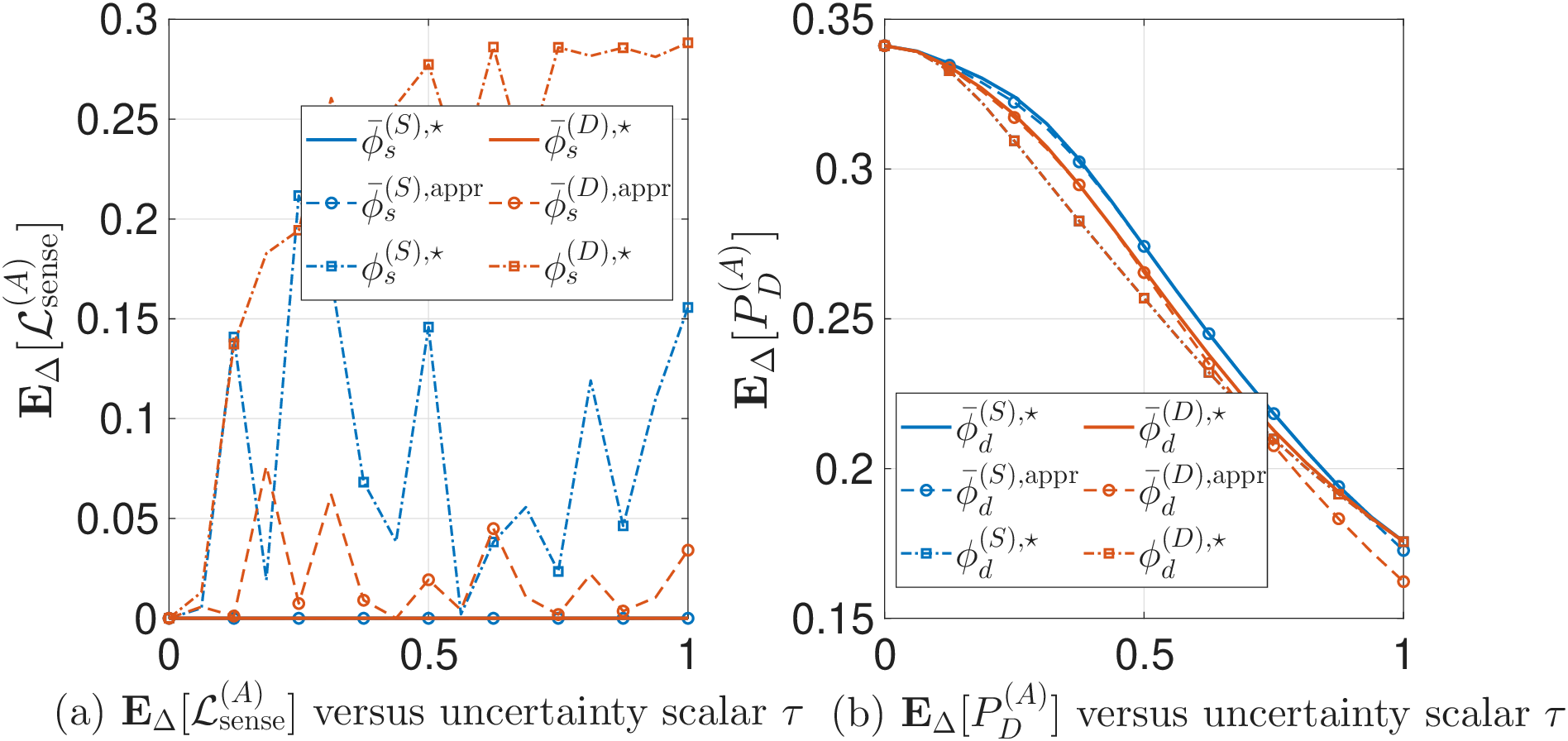}
\caption{Performance of robust range--angle estimation and detection under correlated
Gaussian target-state perturbations. (a) shows
$\mathbb{E}_{\Delta}[\mathcal{L}_{\mathrm{sense}}^{(A)}]$ and (b)
shows $\mathbb{E}_{\Delta}[P_D^{(A)}]$ versus $\tau$ for
$A\in\{S,D\}$. The detection results are for $P_{\rm FA}=10^{-3}$ and
$800$ Monte Carlo samples.}
   \label{robust}
\end{figure}

Fig.~\ref{robust} evaluates robustness to correlated Gaussian target-state
perturbations, with
$\Delta\boldsymbol\psi=[\Delta r_s,\Delta\theta_s]^\top$ and
$\mathbf{\Sigma}_\psi=\tau^2\mathbf{\Sigma}_{\psi,0}$. The perturbation-free rotations
$\phi_{\ell}^{(A),\star}$, with $\ell\in\{s,d\}$ and
$A\in\{S,D\}$, are optimized at the estimated target state without
accounting for target-state errors and then evaluated under
perturbations. The approximate rotations
$\bar{\phi}_{\ell}^{(A),\mathrm{appr}}$, derived in
Lemmas~\ref{robustsensinglemma} and \ref{robustdetectionlemma}, incorporate
the perturbation model into the rotation selection.
For estimation, define
$
\mathcal{L}_{\mathrm{sense}}^{(A)}(\phi,\Delta)
=
10\log_{10}
\frac{\det\!\big(\mathrm{CRB}^{(A)}(\phi,\Delta)\big)}
{\det\!\big(\mathrm{CRB}^{(A)}
(\bar{\phi}_{s}^{(A),\star},\Delta)\big)}.
$
Figure~\ref{robust}(a) plots
$\mathbb{E}_{\Delta}[\mathcal{L}_{\mathrm{sense}}^{(A)}]$ versus
$\tau$. The three curves for each architecture correspond to
$\bar{\phi}_{s}^{(A),\star}$,
$\bar{\phi}_{s}^{(A),\mathrm{appr}}$, and
$\phi_{s}^{(A),\star}$. For \(\tau\in[0,1]\), the approximations to the robust sensing rotations \(\bar{\phi}_{s}^{(A),\mathrm{appr}}\) remain close to the corresponding sensing-centric robust rotations \(\bar{\phi}_{s}^{(A),\star}\), for \(A\in\{S,D\}\).
The largest average loss in the joint range--angle CRB determinant was found to lie below $10^{-6}$~dB for SS (at $\tau=0.375$) and $0.0763$~dB for DW
(at $\tau=0.1875$). The sensing loss of the perturbation-free rotation \(\phi_s^{(A),\star}\) increases as target-state uncertainty grows, with greater degradation for DW\footnote{The approximation of the robust rotation angle for DW omits the $\delta_r$- and
$\delta_\theta$-dependent terms when selecting the angle. Its plotted loss, however, is evaluated using the full CRB expression.}.
Fig.~\ref{robust}(b) shows the average detection probability,
$\mathbb{E}_{\Delta}[P_D^{(A)}]$, evaluated using
\eqref{robustmisnoncentrality} at the detection-centric robust
rotations $\bar{\phi}_{d}^{(A),\star}$, the approximation of the robust
rotation angle $\bar{\phi}_{d}^{(A),\mathrm{appr}}$, and the perturbation-free rotation angle
$\phi_{d}^{(A),\star}$. For \(\tau\in[0,1]\), the approximations to the robust rotations \(\bar{\phi}_{d}^{(A),\mathrm{appr}}\) remain close to the corresponding detection-centric robust rotations \(\bar{\phi}_{d}^{(A),\star}\), for \(A\in\{S,D\}\). The results also show that detection is less affected by
target-state perturbations than range--angle estimation in the
considered cases; thus, robust rotation-angle selection is more important for estimation, while detection remains comparatively insensitive to the perturbations \footnote{It was found that all Gaussian perturbations stayed within the valid propagation region,
so no samples were removed.}.

\begin{figure}[!t]
\centering
\includegraphics[scale=.42]{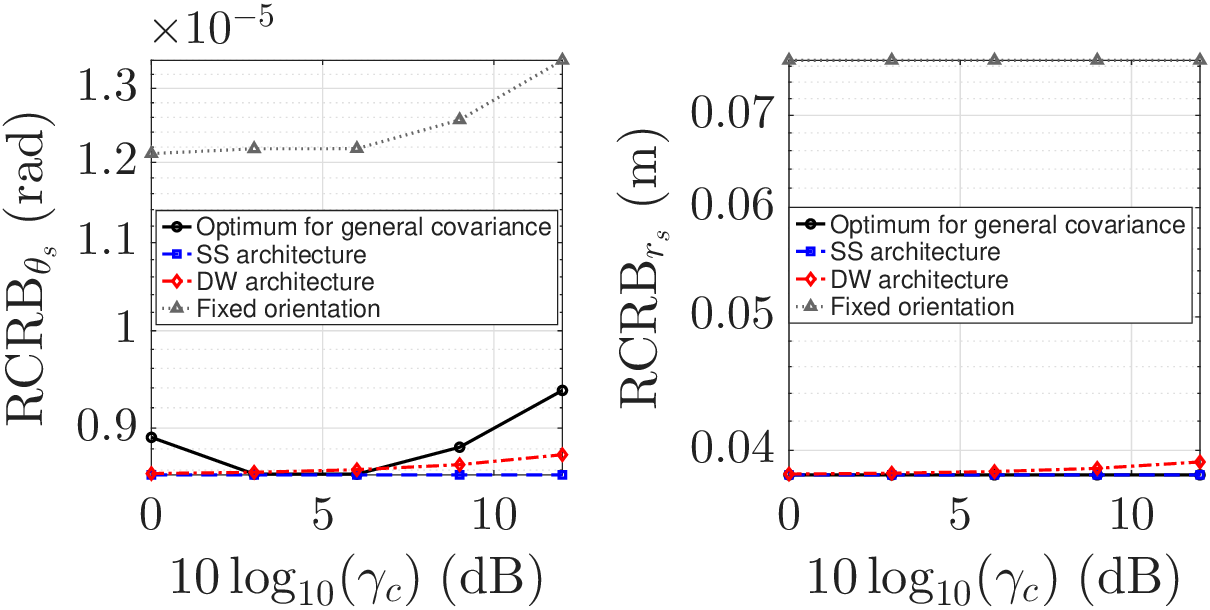}
\caption{Exact finite-array RCRBs versus the minimum required
communication SINR. The left and right subfigures show the angular and range
RCRBs, respectively.}
\label{fig:benchmark}
\end{figure}

Next, in Fig.~\ref{fig:benchmark}, we consider the single-user, single-target special case of the design problem in \cite{Zha26}, which jointly optimized the transmit covariance and rotation angle. The general-covariance result serves as a numerical
reference obtained from the SDR-based covariance-design method and the
one-dimensional rotation search in \cite{Zha26}. SS and DW are evaluated using the same trace--CRB
objective as the general-covariance design, while each is subject to its own coefficient and power-allocation constraints. The curve marked "fixed orientation" evaluates the
general-covariance design with the rotation fixed at $\phi=0$.
Fig.~\ref{fig:benchmark} shows that increasing the minimum required
communication SINR generally worsens both angle and range estimation.
For the considered baseline geometry and over the tested SINR range,
optimized SS remains close to the general-covariance trace optimum in
terms of the trace--CRB objective. Optimized DW follows the same overall
trend but exhibits a larger trace--CRB gap\footnote{Because the
trace--CRB objective is dominated by $\mathrm{CRB}_{r_s}$, the
general-covariance trace optimum need not minimize $\mathrm{CRB}_{\theta_s}$
separately, which explains its larger angle RCRB in the left subfigure.}.
\section{Conclusion}\label{conclusion}
This paper studied how array rotation affects communication, detection, and
range--angle estimation in near-field ISAC with two low-complexity transmit
architectures. The derived CRB decomposition separates the effects of target geometry from
transmit-dependent effects, and explains the different rotation preferences of
the two architectures. The
single-stream design can keep communication and detection compatible,
whereas the dual-waveform split design, in which the user treats the sensing waveform as interference, has a direct communication--detection
tradeoff. The analyses show that rotation provides substantial sensing gains, particularly for range
estimation. We derived compact approximations for the communication-, detection-, and estimation-centric rotation angles, both with and without correlated target-state errors. These approximations closely match the corresponding exact finite-array optimal angles. Detection is less affected by target-state errors than range–angle estimation.
\appendices
\section{Proof of Proposition~\ref{generalCRBprop}}
\label{app:generalcrb}
Let
$
\mathbf u
=
\operatorname{vec}(\mathbf G\mathbf X),
\qquad
\mathbf d_{\iota}
=
\operatorname{vec}
(\dot{\mathbf G}_{\iota}\mathbf X),
\qquad
\iota\in\{r,\theta\}.
$
The mean of the vectorized observation is
$
\boldsymbol{\mu}
=
\beta_s\mathbf u.
$
The derivatives of the mean with respect to the target parameters are
$
\frac{\partial\boldsymbol{\mu}}{\partial r_s}
=
\beta_s\mathbf d_r,
\qquad
\frac{\partial\boldsymbol{\mu}}{\partial\theta_s}
=
\beta_s\mathbf d_\theta.
$
Since
$
\beta_s
=
\Re\{\beta_s\}
+
j\Im\{\beta_s\},
$
the derivatives with respect to the nuisance parameters are
$
\frac{\partial\boldsymbol{\mu}}
{\partial\boldsymbol{\epsilon}}
=
\mathbf u
\begin{bmatrix}
1&j
\end{bmatrix}.
$
For a complex Gaussian observation with real-valued parameters,
the FIM is
$
[\mathbf J_{\boldsymbol\xi}]_{i,j}
=
\frac{2}{\sigma_r^2}
\Re\!\left\{
\frac{\partial\boldsymbol{\mu}^H}{\partial\xi_i}
\frac{\partial\boldsymbol{\mu}}{\partial\xi_j}
\right\}.
$
Using
$
\mathbf X\mathbf X^H=TP_t\mathbf R_x
$
and
$
\operatorname{vec}(\mathbf A)^H
\operatorname{vec}(\mathbf B)
=
\operatorname{tr}(\mathbf A^H\mathbf B),
$
we obtain
$
\mathbf d_{\iota}^H\mathbf d_{\jmath}
=
TP_tT_{\jmath\iota},
$
$
\mathbf d_{\iota}^H\mathbf u
=
TP_tT_{\iota},
$
and
$
\mathbf u^H\mathbf u
=
TP_tT_0.
$
Therefore, the target-parameter block of the FIM is
$
\mathbf J_{\nu\nu}
=
\frac{2|\beta_s|^2TP_t}{\sigma_r^2}
\begin{bmatrix}
T_{rr}&\Re\{T_{\theta r}\}\\
\Re\{T_{\theta r}\}&T_{\theta\theta}
\end{bmatrix}.
$
The cross-information block is
$
\mathbf J_{\nu\boldsymbol{\epsilon}}
=
\frac{2TP_t}{\sigma_r^2}
\begin{bmatrix}
\Re\{\beta_s^*T_r\}
&
-\Im\{\beta_s^*T_r\}
\\
\Re\{\beta_s^*T_\theta\}
&
-\Im\{\beta_s^*T_\theta\}
\end{bmatrix}.
$
Moreover,
$
\mathbf J_{\boldsymbol{\epsilon}\boldsymbol{\epsilon}}
=
\frac{2TP_tT_0}{\sigma_r^2}\mathbf I_2,
$
and hence
$
\mathbf J_{\boldsymbol{\epsilon}\boldsymbol{\epsilon}}^{-1}
=
\frac{\sigma_r^2}{2TP_tT_0}\mathbf I_2.
$
It follows that
$
\mathbf J_{\nu\boldsymbol{\epsilon}}
\mathbf J_{\boldsymbol{\epsilon}\boldsymbol{\epsilon}}^{-1}
\mathbf J_{\nu\boldsymbol{\epsilon}}^\top
=
\frac{2|\beta_s|^2TP_t}{\sigma_r^2T_0}
\begin{bmatrix}
|T_r|^2
&
\Re\{T_rT_\theta^*\}
\\
\Re\{T_rT_\theta^*\}
&
|T_\theta|^2
\end{bmatrix}.
$
Using the Schur complement,
$
\mathbf J
=
\mathbf J_{\nu\nu}
-
\mathbf J_{\nu\boldsymbol{\epsilon}}
\mathbf J_{\boldsymbol{\epsilon}\boldsymbol{\epsilon}}^{-1}
\mathbf J_{\nu\boldsymbol{\epsilon}}^\top.
$
Substitution gives
$
\mathbf J
=
\frac{2|\beta_s|^2TP_t}{\sigma_r^2T_0}
\begin{bmatrix}
T_{rr}T_0-|T_r|^2
&
\Re\{T_{\theta r}T_0-T_rT_\theta^*\}
\\
\Re\{T_{\theta r}T_0-T_rT_\theta^*\}
&
T_{\theta\theta}T_0-|T_\theta|^2
\end{bmatrix}.
$
Therefore,
$
\operatorname{CRB}(\nu)
=
\mathbf J^{-1}
$
which proves Proposition~\ref{generalCRBprop}.
\section{Proof of Lemma \ref{geometryCRBlemma}}\label{geometryCRBproof}
For each element,
$
\frac{\partial b_{s,n}}{\partial \theta_s}=-jk a_n b_{s,n},
\qquad
\frac{\partial b_{s,n}}{\partial r_s}=-jk c_n b_{s,n}.
$
Hence
$
q_r\triangleq \mathbf b_s^H\mathbf b_{s,r}=-jk\sum_{n=1}^N c_n,
\qquad
q_\theta\triangleq \mathbf b_s^H\mathbf b_{s,\theta}=-jk\sum_{n=1}^N a_n,
$
and
$
\|\mathbf b_{s,r}\|^2=k^2S_{c2},
\qquad
\|\mathbf b_{s,\theta}\|^2=k^2S_{a,2},
\qquad
\mathbf b_{s,\theta}^H\mathbf b_{s,r}=k^2S_{a,c}.
$ From the definitions of
$\widetilde{\mathbf b}_r$
and
$\widetilde{\mathbf b}_\theta$,
together with
$\mathbf b_s^H\mathbf b_s=N$,
we obtain
$
\mathbf b_s^H\widetilde{\mathbf b}_r=0,
\qquad
\mathbf b_s^H\widetilde{\mathbf b}_\theta=0.
$
Moreover,
$
\left\|\frac{q_r}{N}\mathbf b_s\right\|^2=\frac{k^2(\sum_{n=1}^N c_n)^2}{N},
\qquad
\left\|\frac{q_\theta}{N}\mathbf b_s\right\|^2=\frac{k^2(\sum_{n=1}^N a_n)^2}{N}.
$
Hence,
$
\|\widetilde{\mathbf b}_r\|^2
=
k^2
\left(
S_{c2}-\frac{(\sum_{n=1}^N c_n)^2}{N}
\right)
=
k^2NV_r.
$
Similarly,
$
\|\widetilde{\mathbf b}_\theta\|^2
=
k^2
\left(
S_{a,2}-\frac{(\sum_{n=1}^N a_n)^2}{N}
\right)
=
k^2NV_\theta,
$
and
$
\widetilde{\mathbf b}_\theta^H\widetilde{\mathbf b}_r
=
k^2
\left(
S_{a,c}-\frac{(\sum_{n=1}^N a_n)(\sum_{n=1}^N c_n)}{N}
\right)
=
k^2NC_{r\theta}.
$
Define
$
\mathbf G
=
\mathbf b_s\mathbf b_s^H.
$
Then
$
\dot{\mathbf G}_r
=
\widetilde{\mathbf b}_r\mathbf b_s^H
+
\mathbf b_s\widetilde{\mathbf b}_r^H,
$
and
$
\dot{\mathbf G}_\theta
=
\widetilde{\mathbf b}_\theta\mathbf b_s^H
+
\mathbf b_s\widetilde{\mathbf b}_\theta^H.
$
First,
$
T_0
=
\operatorname{tr}
(
\mathbf G\mathbf R_x\mathbf G
)
=
Ng_0.
$
Next,
$
T_r
=
\operatorname{tr}
(
\mathbf G\mathbf R_x\dot{\mathbf G}_r
)
=
Ng_1,
$
and similarly
$
T_\theta
=
Ng_2.
$
Expanding
$
T_{rr}
=
\operatorname{tr}
(
\dot{\mathbf G}_r
\mathbf R_x
\dot{\mathbf G}_r
),
$
and using
$
\operatorname{tr}
(
\widetilde{\mathbf b}_r\mathbf b_s^H
)
=
\mathbf b_s^H\widetilde{\mathbf b}_r
=
0,
$
we obtain
$
T_{rr}
=
N
(
k^2g_0V_r+h_{rr}
).
$
Similarly,
$
T_{\theta\theta}
=
N
(
k^2g_0V_\theta+h_{\theta\theta}
).
$
For the cross term,
$
T_{\theta r}
=
\operatorname{tr}
(
\dot{\mathbf G}_\theta
\mathbf R_x
\dot{\mathbf G}_r
)
=
N
(
k^2g_0C_{r\theta}
+
h_{\theta r}
).
$
Substituting the above identities into the definitions in
Proposition~\ref{generalCRBprop} gives
$
m_{11}=N^2A_r,\qquad
m_{22}=N^2A_\theta,\qquad
m_{12}=N^2A_c.
$
Therefore,
$
m_{11}m_{22}-m_{12}^2
=
N^4(A_rA_\theta-A_c^2).
$
Using Proposition~\ref{generalCRBprop} and $T_0=Ng_0$, we obtain
$
\mathbf J
=
\frac{2|\beta_s|^2TP_t}{\sigma_r^2}
\frac{N}{g_0}
\begin{bmatrix}
A_r&A_c\\
A_c&A_\theta
\end{bmatrix}.
$
Inverting $\mathbf J$ gives
$
\operatorname{CRB}(\boldsymbol{\nu})
=
\frac{\sigma_r^2}{2|\beta_s|^2TP_t}
\frac{g_0}{N(A_rA_\theta-A_c^2)}
\begin{bmatrix}
A_\theta&-A_c\\
-A_c&A_r
\end{bmatrix}.
$
Its diagonal entries give the stated range and angle CRBs, and its
determinant gives the stated determinant of the CRB. This completes
the proof.
\section{Proof of Lemma \ref{procrbrankone}}\label{procrbrankoneproof}
We first derive the range–angle CRBs for a rank-one transmit covariance. Let $\mathbf R_x=\mathbf w\mathbf w^H$ and define
$
g\triangleq\mathbf b_s^H\mathbf w,
\qquad
u_r\triangleq\widetilde{\mathbf b}_r^H\mathbf w,
\qquad
u_\theta\triangleq\widetilde{\mathbf b}_\theta^H\mathbf w.
$
Then
$
g_0
=
\mathbf b_s^H\mathbf R_x\mathbf b_s
=
|g|^2,
$
$
g_1
=
\mathbf b_s^H\mathbf R_x\widetilde{\mathbf b}_r
=
g\,u_r^*,
\qquad
g_2
=
g\,u_\theta^*,
$
$
h_{rr}
=
\widetilde{\mathbf b}_r^H\mathbf R_x\widetilde{\mathbf b}_r
=
|u_r|^2,
\qquad
h_{\theta\theta}
=
|u_\theta|^2,
\qquad
h_{\theta r}
=
u_\theta u_r^*.
$
Substituting these into the definitions of $A_r$, $A_\theta$, and $A_c$ in Lemma~\ref{geometryCRBlemma} gives
$
A_r
=
k^2|g|^4V_r,
\qquad
A_\theta
=
k^2|g|^4V_\theta,
\qquad
A_c
=
k^2|g|^4C_{r\theta}.
$
Hence
$
A_rA_\theta-A_c^2
=
k^4|g|^8
\left(
V_rV_\theta-C_{r\theta}^2
\right).
$
Using Lemma~\ref{geometryCRBlemma} and $g_0=|g|^2$ yields $\operatorname{CRB}_{r_s}$ and $\operatorname{CRB}_{\theta_s}$.
\section{Proof of Corollary \ref{procrbrankoneproofedition}}\label{procrbrankoneproofeditionproof}
For any feasible $\mathbf w$, write
$
\mathbf w=\mathbf w_{\parallel}+\mathbf w_{\perp},
$
where
$
\mathbf w_{\parallel}
\in\operatorname{span}\{\mathbf b_s(\phi),\mathbf b_u(\phi)\},
\qquad
\mathbf w_{\perp}
\perp\mathbf b_s(\phi),\mathbf b_u(\phi).
$
Then
$
\mathbf b_s^H(\phi)\mathbf w
=
\mathbf b_s^H(\phi)\mathbf w_{\parallel},
\qquad
\mathbf b_u^H(\phi)\mathbf w
=
\mathbf b_u^H(\phi)\mathbf w_{\parallel}.
$
Since $\Gamma_u>0$ and the problem is feasible,
$\mathbf w_{\parallel}\neq\mathbf 0$. Define
$
\bar{\mathbf w}
=
\frac{\mathbf w_{\parallel}}
{\|\mathbf w_{\parallel}\|}.
$
Because $\|\mathbf w_{\parallel}\|\le\|\mathbf w\|=1$, we have
$
|\mathbf b_s^H(\phi)\bar{\mathbf w}|^2
\ge
|\mathbf b_s^H(\phi)\mathbf w|^2
$
and
$
|\mathbf b_u^H(\phi)\bar{\mathbf w}|^2
\ge
|\mathbf b_u^H(\phi)\mathbf w|^2
\ge\Gamma_u.
$
Thus, $\bar{\mathbf w}$ is feasible and performs no worse than
$\mathbf w$. Therefore, an optimal solution exists in
$
\operatorname{span}\{\mathbf b_s(\phi),\mathbf b_u(\phi)\}.
$
\section{Proof of Lemma \ref{crbBlemma}}\label{crbBlemmaproof}
We begin with the rank-two covariance model
$
\mathbf R_x=\mu_u\mathbf v_u\mathbf v_u^H+\mu_s\mathbf v_s\mathbf v_s^H,
$
and define the scalar projections
$
g(\phi)=\mathbf b_s^H\mathbf v_s,\quad
\rho=\mathbf v_u^H\mathbf v_s,\quad
\rho_c=\frac{1}{N}\sum_{n=1}^N c_n b_{u,n}^*b_{s,n},\quad
\rho_a=\frac{1}{N}\sum_{n=1}^N a_n b_{u,n}^*b_{s,n}.
$
From orthogonal projection definitions,
$
\widetilde{\mathbf b}_r=\mathbf b_{s,r}-\frac{q_r}{N}\mathbf b_s,
\qquad
\widetilde{\mathbf b}_\theta=\mathbf b_{s,\theta}-\frac{q_\theta}{N}\mathbf b_s,
$
and using $q_r=-jkN\bar c$, $q_\theta=-jkN\bar a$, we obtain
$
\mathbf v_u^H\widetilde{\mathbf b}_r
=
-jk\sqrt N(\rho_c-\bar c\rho)
=
-jk\sqrt N\,\delta_r,
$
$
\mathbf v_u^H\widetilde{\mathbf b}_\theta
=
-jk\sqrt N\,\delta_\theta.
$
Similarly,
$
\mathbf b_s^H\mathbf R_x\mathbf b_s
=
\mu_s N+\mu_uN|\rho|^2
=
N A_0,
$
which gives
$
T_0=\operatorname{tr}(\mathbf G\mathbf R_x\mathbf G^H)
=
N\mathbf b_s^H\mathbf R_x\mathbf b_s
=
N^2A_0.
$
Next,
$
T_r
=
N\mathbf b_s^H\mathbf R_x\widetilde{\mathbf b}_r
=
N\big(\mu_u\mathbf b_s^H\mathbf v_u\mathbf v_u^H\widetilde{\mathbf b}_r
+
\mu_s\mathbf b_s^H\mathbf v_s\mathbf v_s^H\widetilde{\mathbf b}_r\big).
$
Using $\mathbf b_s^H\mathbf v_s=\sqrt N$ and $\mathbf b_s^H\mathbf v_u=\sqrt N\rho^*$,
$
T_r
=
N\mu_u\sqrt N\rho^*(-jk\sqrt N\delta_r)
=
-jk\mu_uN^2\rho^*\delta_r.
$
Similarly,
$
T_\theta=-jk\mu_uN^2\rho^*\delta_\theta.
$
For second-order terms,
$
T_{rr}
=
N\widetilde{\mathbf b}_r^H\mathbf R_x\widetilde{\mathbf b}_r
+
\|\widetilde{\mathbf b}_r\|^2\mathbf b_s^H\mathbf R_x\mathbf b_s.
$
We compute
$
\widetilde{\mathbf b}_r^H\mathbf R_x\widetilde{\mathbf b}_r
=
\mu_u|\mathbf v_u^H\widetilde{\mathbf b}_r|^2
+
\mu_s|\mathbf v_s^H\widetilde{\mathbf b}_r|^2
=
\mu_uk^2N|\delta_r|^2,
$
and
$
\|\widetilde{\mathbf b}_r\|^2=k^2NV_r.
$
Thus,
$
T_{rr}=k^2N^2(\mu_u|\delta_r|^2+A_0V_r).
$
Similarly,
$
T_{\theta\theta}=k^2N^2(\mu_u|\delta_\theta|^2+A_0V_\theta).
$
For the cross term,
$
T_{\theta r}
=
\operatorname{tr}(\mathbf R_x\dot{\mathbf G}_r\dot{\mathbf G}_\theta)
=
N\widetilde{\mathbf b}_\theta^H\mathbf R_x\widetilde{\mathbf b}_r
+
(\widetilde{\mathbf b}_r^H\widetilde{\mathbf b}_\theta)\mathbf b_s^H\mathbf R_x\mathbf b_s.
$
Using
$
\widetilde{\mathbf b}_\theta^H\mathbf R_x\widetilde{\mathbf b}_r
=
\mu_uk^2N\delta_\theta^*\delta_r,
\quad
\widetilde{\mathbf b}_r^H\widetilde{\mathbf b}_\theta=k^2NC_{r\theta},
$
we obtain
$
T_{\theta r}=k^2N^2(\mu_u\delta_\theta^*\delta_r+A_0C_{r\theta}).
$
Substituting these expressions into the Schur complement definitions of
$m_{11}, m_{22}, m_{12}$ and collecting terms yields
$
m_{11}^{(D)}
=
k^2N^4\big(A_0^2V_r+\mu_u\mu_s|\delta_r|^2\big),
$
$
m_{22}^{(D)}
=
k^2N^4\big(A_0^2V_\theta+\mu_u\mu_s|\delta_\theta|^2\big),
$
$
m_{12}^{(D)}
=
k^2N^4\big(A_0^2C_{r\theta}+\mu_u\mu_s\Re\{\delta_\theta^*\delta_r\}\big),
$
where the identity $A_0-\mu_u|\rho|^2=\mu_s$ is used to cancel mixed terms. Finally, substituting into the CRB expression
$\operatorname{CRB}(\nu)=\mathbf J^{-1}$
yields the stated closed-form range–angle CRBs.
\section{Proof of Proposition~\ref{prop:derivativediversity}}
\label{prop:derivativediversityproof}
For SS, substituting the rank-one covariance into the definitions of
$T_0$ and $m_{ij}$ in Proposition~\ref{generalCRBprop} gives
$
T_0^{(S)}=N|g^{(S)}|^2,
\qquad
\begin{bmatrix}
m_{11}^{(S)}&m_{12}^{(S)}\\
m_{12}^{(S)}&m_{22}^{(S)}
\end{bmatrix}
=
k^2N^2|g^{(S)}|^4\mathbf V_s.
$
Therefore, Proposition~\ref{generalCRBprop} yields
$
\mathbf J^{(S)}
=
\frac{\kappa}{T_0^{(S)}}
\begin{bmatrix}
m_{11}^{(S)}&m_{12}^{(S)}\\
m_{12}^{(S)}&m_{22}^{(S)}
\end{bmatrix}
=
\kappa k^2N|g^{(S)}|^2\mathbf V_s.
$
For DW, the corresponding substitutions give
$
T_0^{(D)}=N^2A_0,
\qquad
\begin{bmatrix}
m_{11}^{(D)}&m_{12}^{(D)}\\
m_{12}^{(D)}&m_{22}^{(D)}
\end{bmatrix}
=
k^2N^4
\left(
A_0^2\mathbf V_s+\mu_u\mu_s\mathbf D_\delta
\right).
$
Hence,
$
\begin{aligned}
\mathbf J^{(D)}
=
\frac{\kappa}{T_0^{(D)}}
\begin{bmatrix}
m_{11}^{(D)}&m_{12}^{(D)}\\
m_{12}^{(D)}&m_{22}^{(D)}
\end{bmatrix}=
\kappa k^2N^2
(
A_0\mathbf V_s+
\frac{\mu_u\mu_s}{A_0}\mathbf D_\delta
).
\end{aligned}
$
Finally, for any $\mathbf x=[x_1,x_2]^\top\in\mathbb R^2$,
$
\mathbf x^\top\mathbf D_\delta\mathbf x
=
\left|x_1\delta_r+x_2\delta_\theta\right|^2\geq0.
$
Thus, $\mathbf D_\delta\succeq0$. It is zero if and only if
$\delta_r=\delta_\theta=0$. Therefore, for $0<\mu_u<1$ and $A_0>0$,
the additional DW term is positive semidefinite and vanishes if and only
if $\delta_r=\delta_\theta=0$. This proves the proposition.
\section{Proof of Lemma~\ref{thm:detection-optimality}}
\label{app:detection-optimality}
We fix $\phi$ and suppress its argument.
Since $T_0=\operatorname{tr}(\mathbf G\mathbf R_x\mathbf G^H)
=N^2\mathbf v_s^H\mathbf R_x\mathbf v_s$ and $Q_1(\cdot,b)$ is strictly
increasing in its first argument, \eqref{eq:det-cov-problem} is
equivalent to
\begin{align}
\max_{\mathbf R_x\succeq\mathbf 0}\;
\mathbf v_s^H\mathbf R_x\mathbf v_s
\quad\text{s.t.}\quad
\operatorname{tr}(\mathbf R_x)=1,\;\;
\mathbf v_u^H\mathbf R_x\mathbf v_u\geq q .
\label{eq:det-linear}
\end{align}
Problem \eqref{eq:det-linear} is feasible for every $q\le1$, since
$\mathbf v_u\mathbf v_u^H$ satisfies both constraints. Its feasible set is
compact, because a positive semidefinite matrix with unit trace has all
eigenvalues in $[0,1]$. Hence its linear objective attains a maximum.
Moreover, \eqref{eq:det-linear} is a complex-valued SDP with a linear
objective and only two real linear constraints. By the rank-reduction
result for complex SDPs in \cite{Huang10,Pataki98},
there exists an optimal solution of rank $r$ satisfying $r^2\leq 2$,
i.e., $r=1$. Thus, an optimal solution can be written as
$\mathbf R_x^\star=\mathbf w\mathbf w^H$ with $\|\mathbf w\|=1$.
Let $\mathcal P\triangleq\operatorname{span}\{\mathbf v_u,\mathbf v_s\}
=\operatorname{span}\{\mathbf b_u,\mathbf b_s\}$ with orthogonal projector
$\mathbf P_{\mathcal P}$, and let $p\triangleq\|\mathbf P_{\mathcal P}\mathbf w\|\le1$.
Since $\mathbf v_u,\mathbf v_s\in\mathcal P$, we have
$\mathbf v_i^H\mathbf P_{\mathcal P}\mathbf w=\mathbf v_i^H\mathbf w$ for
$i\in\{u,s\}$. In particular, $p>0$, since otherwise
$\mathbf v_u^H\mathbf w=0$, contradicting the QoS constraint with $q>0$.
Hence $\bar{\mathbf w}\triangleq\mathbf P_{\mathcal P}\mathbf w/p$ is unit
norm and satisfies
$|\mathbf v_i^H\bar{\mathbf w}|^2=|\mathbf v_i^H\mathbf w|^2/p^2
\geq|\mathbf v_i^H\mathbf w|^2$, $i\in\{u,s\}$.
Thus $\bar{\mathbf w}\bar{\mathbf w}^H$ is feasible and achieves an
objective value no smaller than that of $\mathbf w\mathbf w^H$, so it is
also optimal. Since $\bar{\mathbf w}\in\mathcal P$, it can be written as
the normalized combination \eqref{beamA} of $\mathbf v_u$ and $\mathbf v_s$
for some $(c_u,c_s)$. Its QoS constraint satisfies
$|c_u+c_s\rho|^2/Z^{(S)}\geq q$. This completes the proof.
\section{Proof of Corollary~\ref{cor:detection-dw}}
\label{proof:detection-dw}
From \eqref{SINRB},
$\mathrm{SINR}_u^{(D)}\ge\gamma_{u,\min}$ gives
$\mu_u\tilde\Gamma_u\ge\gamma_{u,\min}
\big(1+(1-\mu_u)\tilde\Gamma_u|\rho|^2\big)\ge\gamma_{u,\min}$,
i.e., $\mu_u\ge q$. Hence the DW covariance
$\mathbf R_x^{(D)}=\mu_u\mathbf v_u\mathbf v_u^H+\mu_s\mathbf v_s\mathbf v_s^H$
satisfies $\operatorname{tr}(\mathbf R_x^{(D)})=1$ and
$\mathbf v_u^H\mathbf R_x^{(D)}\mathbf v_u=\mu_u+\mu_s|\rho|^2\ge\mu_u\ge q$,
so it is feasible for \eqref{eq:det-cov-problem}. By
Lemma~\ref{thm:detection-optimality}, an SS beamformer
$\mathbf w^{(S)}$ attains the optimum of \eqref{eq:det-cov-problem}, so
$P_D^{(S)}\ge P_D^{(D)}$. Finally, \eqref{SINRA} can be written as
$\mathrm{SNR}_u^{(S)}=\tilde\Gamma_u|\mathbf v_u^H\mathbf w^{(S)}|^2
\ge\tilde\Gamma_u q=\gamma_{u,\min}$, so this SS design meets the same
SINR requirement.
\section{Proof of Corollary~\ref{cor:exactoverlap}}
\label{app:exactoverlap}
For Case 1, let
$
\tilde\phi=(\theta_u+\theta_s)/2.
$
Then
$
\theta_s-\tilde\phi=(\theta_s-\theta_u)/2,
\qquad
\theta_u-\tilde\phi=-(\theta_s-\theta_u)/2.
$
Since
$
r_u=r_s,
$
we obtain
$
r_n(\theta_s,r_u,\tilde\phi)=r_n(\theta_u,r_u,\tilde\phi)
$
for all $n$, hence
$
\mathbf b_s(\tilde\phi)=\mathbf b_u(\tilde\phi)
$
and
$
\mathbf v_s(\tilde\phi)=\mathbf v_u(\tilde\phi).
$
Therefore
$
\mathbf w^{(S)}=\frac{(c_u+c_s)\mathbf v_u}{|c_u+c_s|}=e^{j\psi}\mathbf v_u
$
for some phase $\psi$, so
$
|\mathbf v_u^H\mathbf w^{(S)}|^2=1.
$
Thus
$
\mathrm{SNR}_u^{(S)}(\tilde\phi)=\frac{P_t|\beta_u|^2N}{\sigma_u^2},
$
which is the largest achievable value for any unit-norm beamformer. Hence
$
\tilde\phi
$
is globally optimal.
For Case 2, let
$
\tilde\phi=\theta_0.
$
Then
$
r_n(\theta_0,r,\tilde\phi)
=
\sqrt{r^2+\chi_n^2d^2-2r\chi_nd}
=
|r-\chi_nd|.
$
Since
$
r>(N-1)d/2\ge |\chi_n|d
$
for all $n$, we obtain
$
r_n-r=-\chi_nd,
$
which is independent of $r$. Hence
$
\mathbf b(\theta_0,r_u,\tilde\phi)=\mathbf b(\theta_0,r_s,\tilde\phi),
$
so
$
\mathbf v_u(\tilde\phi)=\mathbf v_s(\tilde\phi).
$
The same argument as in Case 1 then shows that
$
\tilde\phi
$
is globally optimal.
For Case 3(a), Case 1 already gives
$
\mathbf b_u(\tilde\phi)=\mathbf b_s(\tilde\phi)
$
with
$
\tilde\phi=(\theta_u+\theta_s)/2.
$
For Case 3(b), Case 2 gives
$
\mathbf b_u(\tilde\phi)=\mathbf b_s(\tilde\phi)
$
with
$
\tilde\phi=\theta_0.
$
Hence in both cases
$
|\rho(\tilde\phi)|=1.
$
Since
$
|\rho(\phi)|\le 1
$
by Cauchy--Schwarz, these choices globally maximize
$
|\rho(\phi)|^2,
$
and therefore also
$
P_D^{(D)}(\phi\mid\beta_s).
$
\section{Proof of Corollary~\ref{cor:secondorderest}}
\label{app:secondorderest}
For Case 1, both architectures satisfy
$
\mathbf R_x=\mathbf v_s\mathbf v_s^H,
$
so their sensing-centric objectives coincide. under the second-order approximation,
$
V_r(\phi)V_\theta(\phi)-C_{r\theta}^2(\phi)\propto \sin^6(\theta_s-\phi),
$
which is maximized at broadside. Also,
$
\mathrm{CRB}_{\theta_s}\propto \sin^{-2}(\theta_s-\phi),
$
and
$
\mathrm{CRB}_{r_s}\propto
\frac{a+b\cos^2(\theta_s-\phi)}{\sin^4(\theta_s-\phi)},
\qquad
a>0,\;b>0,
$
so both CRBs are minimized at broadside, i.e.,
$
\phi\equiv \theta_s-\frac{\pi}{2}.
$
For Case 2, co-location gives
$
\mathbf b_u(\phi)=\mathbf b_s(\phi)
$
for all $\phi$, hence
$
\rho(\phi)=1
$
for all
$
\phi\in\Phi.
$
Under SS,
$
\mathbf w^{(S)}=\frac{(c_u+c_s)\mathbf v_s}{|c_u+c_s|}=e^{j\psi}\mathbf v_s,
$
and under DW,
$
\mathbf R_x
=
\mu_u\mathbf v_s\mathbf v_s^H+\mu_s\mathbf v_s\mathbf v_s^H
=
\mathbf v_s\mathbf v_s^H.
$
Thus the effective beam is target-directed in both cases, so the communication and detection terms are constant in $\phi$ and the joint objective reduces to the sensing objective. The broadside conclusion then follows from Case 1.
\section{Proof of Lemma~\ref{robustsensinglemma}}
\label{app:robustsensing}
For a fixed $\Delta$, the transmit waveform is designed using
$(\hat r_s,\hat\theta_s)$ and is held fixed when differentiating with
respect to the true target state
$(r_s,\theta_s)=(\hat r_s+\Delta r_s,\hat\theta_s+\Delta\theta_s)$. Therefore, the
Fisher-information derivation is identical to that in
Section~\ref{sensingestimation}, with the target geometry evaluated at
$
(r_s,\theta_s)
=
(\hat r_s+\Delta r_s,\hat\theta_s+\Delta\theta_s).
$
Thus, the CRB expression in Lemma~\ref{procrbrankone} for SS and
Lemma~\ref{crbBlemma} for DW retain their baseline forms with the mismatch-dependent quantities derived below.
For notational consistency, define
$
\hat{\mathbf b}_s(\phi)
\triangleq
\sqrt N\,\hat{\mathbf v}_s(\phi),
\qquad
\hat b_{s,n}(\phi)
\triangleq
[\hat{\mathbf b}_s(\phi)]_n,
$
and
$
b_{s,n}(\phi,\Delta)
\triangleq
[\sqrt N\,\mathbf v_s(\phi,\Delta)]_n.
$
For $i\in\{r,\theta\}$, let
$
q_{r,n}=c_n,
\qquad
q_{\theta,n}=a_n,
\qquad
\bar q_i=\frac{1}{N}\sum_{n=1}^{N}q_{i,n},
$
where the geometry quantities are evaluated at the perturbed target state.
Define
$
\rho_i(\phi,\Delta)
=
\frac{1}{N}
\sum_{n=1}^{N}
q_{i,n}
b_{u,n}^*(\phi)b_{s,n}(\phi,\Delta),
$
and
$
\eta_i(\phi,\Delta)
=
\frac{1}{N}
\sum_{n=1}^{N}
q_{i,n}
\hat b_{s,n}^*(\phi)b_{s,n}(\phi,\Delta).
$
The corresponding residual derivative terms are
$
\delta_i^{(u)}
=
\rho_i-\bar q_i\rho_\Delta,
\qquad
\delta_i^{(\hat s)}
=
\eta_i-\bar q_i\eta,
$
and therefore
$
\widetilde{\delta}_i
=
\eta\,\delta_i^{(u)}
-
\rho_\Delta\,\delta_i^{(\hat s)}.
$
Thus, in the DW CRB expressions under target-state mismatch,
replace the baseline terms $\delta_i$ by
$\widetilde{\delta}_i$, for $i\in\{r,\theta\}$.
Projecting the target steering vector at the perturbed state onto the
SS beamformer designed using $(\hat r_s,\hat\theta_s)$ gives
$
\mathbf b_s^H(\phi,\Delta)\mathbf w^{(S)}(\phi)
=g^{(S)}(\phi,\Delta).
$ Similarly, the DW target
illumination quantity becomes
$
A_0(\phi,\Delta)
=
\mu_u|\rho_\Delta(\phi)|^2
+
\mu_s|\eta(\phi,\Delta)|^2.
$
Under the second-order approximation,
$
V_r(\phi,\Delta)V_\theta(\phi,\Delta)
-
C_{r\theta}^2(\phi,\Delta)
=
K(\Delta)
\sin^6(\theta_s-\phi),
$
where $K(\Delta)>0$ is independent of $\phi$. Since
$
\theta_s=\hat\theta_s+\Delta\theta_s,
$
the expected geometry contribution is
$
\mathbb E_\Delta
\left[
6\log|\sin(\theta_s-\phi)|
\right]
=
\mathcal G_6(\hat\theta_s-\phi).
$
For SS, the remaining rotation-dependent term in
$-\log\det(\operatorname{CRB}^{(S)})$ is
$
\log|g^{(S)}(\phi,\Delta)|^4.
$
For DW, after neglecting the perturbed derivative-diversity terms, it is
$
\log A_0(\phi,\Delta)^2.
$
Taking expectations gives the two simplified robust sensing objectives stated
in Lemma~\ref{robustsensinglemma}.
\section{Proof of Lemma~\ref{robustdetectionlemma}}
\label{app:robustdetection}
For fixed $P_{\rm FA}$, the detection probability is monotone increasing in
the mismatched-GLRT noncentrality
$\lambda_{\rm det}^{\rm mis}(\phi,\Delta)$. Using the nominal SS and DW
transmit designs in \eqref{robustmisnoncentrality}, and dropping positive
factors independent of $\phi$, the conditional noncentrality factors become
$
\lambda_{\rm det}^{(S)}(\phi,\Delta)
\propto
|\eta(\phi,\Delta)|^2
|g^{(S)}(\phi,\Delta)|^2
$
for SS, and
$
\lambda_{\rm det}^{(D)}(\phi,\Delta)
\propto
|\eta(\phi,\Delta)|^2
\frac{
\left|
\mu_u\hat\rho^*(\phi)\rho_\Delta(\phi)
+
\mu_s\eta(\phi,\Delta)
\right|^2
}{
\mu_s+\mu_u|\hat\rho(\phi)|^2
}
$
for DW.
For sufficiently small perturbations, a first-order expansion of the
target steering vector around the nominal state gives
$
\mathbf b_s(\phi,\Delta)
\approx
\hat{\mathbf b}_s(\phi)
+
\hat{\mathbf b}_{s,r}(\phi)\Delta r_s
+
\hat{\mathbf b}_{s,\theta}(\phi)\Delta\theta_s.
$
After removing the component parallel to
$\hat{\mathbf b}_s(\phi)$, the corresponding coherence satisfies
$
|\eta(\phi,\Delta)|^2
\approx
1-k^2\Big(
\hat V_r(\phi)\Delta r_s^2
+
\hat V_\theta(\phi)\Delta\theta_s^2
+
2\hat C_{r\theta}(\phi)
\Delta r_s\Delta\theta_s
\Big).
$
Taking expectations yields
$
\mathbb E_\Delta[|\eta(\phi,\Delta)|^2]
\approx
1-\upsilon(\phi),
$
where
$
\upsilon(\phi)
=
k^2\Big(
\hat V_r(\phi)\sigma_{\Delta r}^2
+
\hat V_\theta(\phi)\sigma_{\Delta\theta}^2
+
2\hat C_{r\theta}(\phi)\sigma_{r\theta}
\Big).
$
Replacing the remaining state-dependent gain terms by their nominal values
gives
$
\mathcal I_{\rm det}^{(S)}(\phi)
=
|\hat g^{(S)}(\phi)|^2,
\qquad
\mathcal I_{\rm det}^{(D)}(\phi)
=
\mu_s+\mu_u|\hat\rho(\phi)|^2.
$
Thus, after neglecting higher-order uncertainty terms, the expected
noncentrality factors are proportional to
$
\mathcal I_{\rm det}^{(A)}(\phi)
\big(1-\upsilon(\phi)\big),
\qquad
A\in\{S,D\}.
$
Accordingly, the rotation is optimized only over
$\Phi_{\mathrm{valid}}$. For $\upsilon(\phi)\geq1$, the factor
$1-\upsilon(\phi)$ is nonpositive, so those angles are excluded from
$\Phi_{\mathrm{valid}}$.
When $\mathbf\Sigma_\psi=\mathbf0$, the target-state error is zero, so
$\eta=1$, $\rho_\Delta=\hat\rho$, and $\upsilon=0$. The robust
detection objective then coincides with the baseline detection
objective evaluated at $(\hat r_s,\hat\theta_s)$. Hence the robust
rotation reduces to the detection-centric rotation designed for that
estimated state without target-state error.
\bibliographystyle{IEEEtran}
\bibliography{ref}

@inproceedings{Liu22,
  author    = {Fan Liu and Ya-Feng Liu and Christos Masouros and Ang Li and Yonina C. Eldar},
  title     = {A Joint Radar-Communication Precoding Design Based on Cram{\'e}r-Rao Bound Optimization},
  booktitle = {2022 IEEE Radar Conference (RadarConf22)},
  year      = {2022},
  pages     = {1--6}
}

@misc{Wan26o,
      title={Sensing-Aided Secure Multicast in Two-Level Rotatable Antenna-Enabled ISAC Systems: Modeling and Optimization}, 
      author={Zequan Wang and Liang Yin and Hao Xu and Yunan Sun and Yitong Liu and Hongwen Yang},
      year={2026},
      eprint={2605.08718},
      archivePrefix={arXiv},
      primaryClass={cs.IT},
      url={https://arxiv.org/abs/2605.08718}, 
}

@article{Ngu24b,
  author  = {Nguyen, Nhan Thanh and
             Nguyen, Van-Dinh and
             Nguyen, Hieu V. and
             Ngo, Hien Quoc and
             Swindlehurst, A. Lee and
             Juntti, Markku J.},
  title   = {Performance Analysis and Power Allocation for Massive {MIMO} {ISAC} Systems},
  journal = {IEEE Transactions on Signal Processing},
  volume  = {73},
  pages   = {1691--1707},
  year    = {2025},
  doi     = {10.1109/TSP.2025.3554012}
}

@article{Shi26b,
  author  = {Shi, Jun-Jie and
             Xu, Wei-Yang and
             Pan, Cun-Hua and
             Ngo, Hien Quoc},
  title   = {Active Eavesdropping in Cell-Free {mMIMO} {ISAC} Systems},
  journal = {IEEE Transactions on Wireless Communications},
  volume  = {25},
  pages   = {23781--23796},
  year    = {2026},
  doi     = {10.1109/TWC.2026.3735511}
}

@article{Wu26f,
  author  = {Wu, Jilong and
             Li, Jiamin and
             Zhu, Pengcheng and
             Zhu, Lei},
  title   = {Resource Allocation for Cooperative {ISAC} in {NAFD} Cell-Free {mMIMO} Systems With Jamming},
  journal = {IEEE Transactions on Vehicular Technology},
  year    = {2026},
  doi     = {10.1109/TVT.2026.3734080}
}

@article{Ouy23b,
  author  = {Ouyang, Chong-Jun and
             Liu, Yuan-Wei and
             Zhang, Xing-Qi},
  title   = {Revealing the Impact of Beamforming in {ISAC}},
  journal = {IEEE Wireless Communications Letters},
  volume  = {13},
  number  = {2},
  pages   = {362--366},
  year    = {2024},
  doi     = {10.1109/LWC.2023.3329353}
}

@article{Zha23,
  author  = {Zhao, Boqun and
             Ouyang, Chongjun and
             Liu, Yuanwei and
             Zhang, Xingqi and
             Poor, H. Vincent},
  title   = {Modeling and Analysis of Near-Field {ISAC}},
  journal = {IEEE Journal of Selected Topics in Signal Processing},
  volume  = {18},
  number  = {4},
  pages   = {678--693},
  year    = {2024},
  doi     = {10.1109/JSTSP.2024.3386054}
}

@article{Huang10,
  author  = {Y. Huang and D. P. Palomar},
  title   = {Rank-Constrained Separable Semidefinite Programming With Applications to Optimal Beamforming},
  journal = {IEEE Trans. Signal Process.},
  volume  = {58}, number = {2}, pages = {664--678}, year = {2010}}

@article{Pataki98,
  author  = {G. Pataki},
  title   = {On the Rank of Extreme Matrices in Semidefinite Programs and the Multiplicity of Optimal Eigenvalues},
  journal = {Math. Oper. Res.},
  volume  = {23}, number = {2}, pages = {339--358}, year = {1998}}

@ARTICLE{selvan,
  author={Selvan, Krishnasamy T. and Janaswamy, Ramakrishna},
  journal={IEEE Antennas and Propagation Magazine}, 
  title={Fraunhofer and Fresnel Distances: Unified derivation for aperture antennas}, 
  year={2017},
  volume={59},
  number={4},
  pages={12-15},
  doi={10.1109/MAP.2017.2706648}}

@ARTICLE{CrameRaoBoundOptimizationforJoint,
  author={Liu, Fan and Liu, Ya-Feng and Li, Ang and Masouros, Christos and Eldar, Yonina C.},
  journal={IEEE Transactions on Signal Processing}, 
  title={Cramér-Rao Bound Optimization for Joint Radar-Communication Beamforming}, 
  year={2022},
  volume={70},
  number={},
  pages={240-253},
  doi={10.1109/TSP.2021.3135692}}

@article{Lu24,
  author  = {Zhizheng Lu and Yu Han and Shi Jin and Michail Matthaiou},
  title   = {Near-Field Localization and Channel Reconstruction for ELAA Systems},
  journal = {IEEE Transactions on Wireless Communications},
  year    = {2024},
  pages   = {6938--6953}
}

@article{Lu21b,
  author  = {Haiquan Lu and Yong Zeng},
  title   = {Communicating With Extremely Large-Scale Array/Surface: Unified Modeling and Performance Analysis},
  journal = {IEEE Transactions on Wireless Communications},
  year    = {2021},
  pages   = {4039--4053}
}

@article{Wan23,
  author  = {Huizhi Wang and Zhiqiang Xiao and Yong Zeng},
  title   = {Cram{\'e}r-Rao Bounds for Near-Field Sensing With Extremely Large-Scale MIMO},
  journal = {IEEE Transactions on Signal Processing},
  year    = {2023},
  pages   = {701--717}
}

@article{Wan23c,
  author  = {Zhao-Jie Wang and Xidong Mu and Yuanwei Liu},
  title   = {Near-Field Integrated Sensing and Communications},
  journal = {IEEE Communications Letters},
  year    = {2023},
  pages   = {2048--2052}
}

@article{Hua24,
  author  = {Haocheng Hua and Jie Xu and Rui Zhang},
  title   = {Near-Field Integrated Sensing and Communication With Extremely Large-Scale Antenna Array},
  journal = {IEEE Transactions on Wireless Communications},
  year    = {2024},
  pages   = {9962--9977}
}

@article{Li23b,
  author  = {Haochen Li and Zhao-Jie Wang and Xidong Mu and Zhiwen Pan and Yuanwei Liu},
  title   = {Near-Field Integrated Sensing, Positioning, and Communication: A Downlink and Uplink Framework},
  journal = {IEEE Journal on Selected Areas in Communications},
  year    = {2023},
  pages   = {2196--2212}
}

@article{Luo25,
  author  = {Jie Luo and Jiancun Fan and Yuanwei Liu},
  title   = {Beam Focusing for Near-Field Integrated Sensing and Communications With Hybrid Analog/Digital Architecture},
  journal = {IEEE Transactions on Wireless Communications},
  year    = {2025},
  pages   = {8162--8177}
}

@article{Ray26,
  author  = {Ashirwad Ray and S. Majhi and Harshit Rajput and Marco Di Renzo},
  title   = {ISAC for Near-Field and Extremely Large-Scale Uniform Planar Array Systems},
  journal = {IEEE Transactions on Communications},
  year    = {2026},
  pages   = {1497--1512}
}

@article{Hua22,
  author  = {Haocheng Hua and T. Han and Jie Xu},
  title   = {MIMO Integrated Sensing and Communication: CRB-Rate Tradeoff},
  journal = {IEEE Transactions on Wireless Communications},
  year    = {2022},
  pages   = {2839--2854}
}

@article{Gal24,
  author  = {D. Galappaththige and Shayan Zargari and C. Tellambura and Geoffrey Ye Li},
  title   = {Near-Field ISAC: Beamforming for Multi-Target Detection},
  journal = {IEEE Wireless Communications Letters},
  year    = {2024},
  pages   = {1938--1942}
}

@article{Zho25,
  author  = {Chao Zhou and Changsheng You and Beixiong Zheng and Xiaodan Shao and Rui Zhang},
  title   = {Rotatable Antennas for Integrated Sensing and Communications},
  journal = {IEEE Wireless Communications Letters},
  year    = {2025},
  pages   = {2838--2842}
}

@inproceedings{Jay25,
  author    = {M. A. Jayasinghe and J. K. Dassanayake and Gayan Amarasuriya},
  title     = {Trade-Off Between Probability of Detection and Achievable Rate in Near-Field ISAC Systems},
  booktitle = {2025 IEEE International Conference on Communications (ICC)},
  year      = {2025},
  pages     = {1766--1771}
}

@article{Zha26,
  author  = {Yunpu Zhang and H. C. So and Dusit Niyato and Christos Masouros},
  title   = {Rotatable Antennas for Near-Field Integrated Sensing and Communication},
  journal = {IEEE Transactions on Wireless Communications},
  year    = {2026},
  pages   = {10986--11001}
}

@article{Sun25,
  author  = {Yunan Sun and Hao Xu and Chongjun Ouyang and Hongwen Yang},
  title   = {Rotatable and Movable Antenna-Enabled Near-Field Integrated Sensing and Communication},
  journal = {IEEE Internet of Things Journal},
  year    = {2025},
  pages   = {45119--45132}
}

@article{Wan24,
  author  = {Zhao-Jie Wang and Xidong Mu and Yuanwei Liu},
  title   = {Performance Analysis of Near-Field Sensing in Wideband MIMO Systems},
  journal = {IEEE Transactions on Wireless Communications},
  year    = {2024},
  pages   = {8236--8251}
}

@article{Jia24c,
  author  = {Hao Jiang and Zhao-Jie Wang and Yuanwei Liu and Arumugam Nallanathan},
  title   = {Cram{\'e}r--Rao Bound Optimization for Near-Field Sensing With Continuous-Aperture Arrays},
  journal = {IEEE Transactions on Wireless Communications},
  year    = {2024},
  pages   = {7032--7047}
}

@article{Wan25,
  author  = {Zequan Wang and Liang Yin and Zimeng Lei and Yitong Liu and Yunan Sun and Hongwen Yang},
  title   = {Rotatable Array-Aided Hybrid Beamforming for Integrated Sensing and Communication},
  journal = {IEEE Internet of Things Journal},
  year    = {2025},
  pages   = {14364--14377}
}

@article{Qu25,
  author  = {Kaiqian Qu and Haojin Li and Chen Sun and Wenqi Zhang and Shuaishuai Guo and Haijun Zhang},
  title   = {Rotatable Array-Enabled Multi-BS Cooperative ISAC Transmit Beampattern Design},
  journal = {IEEE Transactions on Vehicular Technology},
  year    = {2025},
  pages   = {14775--14780}
}

@article{Qu23,
  author  = {Kaiqian Qu and Shuaishuai Guo and Nasir Saeed and Jia Ye},
  title   = {Near-Field Integrated Sensing and Communication: Performance Analysis and Beamforming Design},
  journal = {IEEE Open Journal of the Communications Society},
  year    = {2023},
  pages   = {6353--6366}
}
\end{document}